\documentclass{tlp}

\usepackage{amsmath}
\usepackage{amssymb}

\let\emptyset=\varnothing

\newcommand{\lra}{\mathop{\Longrightarrow}\limits}
\newcommand{\la}{\leftarrow}
\newcommand{\ie}{\emph{i.e.}}
\newcommand{\eg}{\emph{e.g.}}
\newcommand{\Var}{\mathit{Var}}

\newcommand{\query}[2]{\langle{#1}\,|\,{#2}\rangle}
\newcommand{\restrict}[2]{#1_{#2}}
\newcommand{\rel}[1]{\mathit{rel}({#1})}
\newcommand{\arity}[1]{\mathit{arity}({#1})}
\newcommand{\sat}[2]{\mathit{sat}({#1},{#2})}
\newcommand{\Set}[1]{\mathit{Set}({#1})}

\newcommand{\Q}{\mathbb{Q}}

\newtheorem{theorem}{Theorem}[section]
\newtheorem{corollary}[theorem]{Corollary}

\newtheorem{lemma}[theorem]{Lemma}
\newtheorem{definition}[theorem]{Definition}
\newtheorem{example}[theorem]{Example}

\title[Non-Termination of CLP]
{A Non-Termination Criterion for Binary Constraint Logic Programs}
\author[\'E.~Payet and F.~Mesnard]
{\'ETIENNE PAYET and FRED MESNARD\\
  IREMIA -  LIM - universit\'e de la R\'eunion, France\\
  \email{\{epayet,fred\}@univ-reunion.fr} }

\submitted{9 July 2007}
\revised{6 August 2008}
\accepted{9 January 2009}

\begin{document}

\maketitle

\noindent
\textbf{Note:} this paper has been accepted for publication in
Theory and Practice of Logic Programming (TPLP).\\

\begin{abstract}
On  the one hand, termination analysis of logic programs 
is now a fairly established research topic
within the logic programming community. On the other hand,
non-termination analysis seems to remain a much less attractive subject. 
If we divide this line of research into two kinds of approaches: dynamic
versus static analysis, this paper belongs to the latter. It proposes
a criterion for detecting non-terminating atomic queries  with respect to
binary CLP rules, which strictly generalizes our previous works on this subject.
We give a generic operational definition and an implemented
logical form of this criterion. 
Then we show that the logical form is correct and complete with respect 
to the operational definition. 
\end{abstract}



\section{Introduction}\label{section-intro}

On  the one hand, termination analysis of logic programs is
a fairly established research topic
within the logic programming community, see the following surveys:
\cite{SD94,Mesnard03b}.
Various termination analyzers are now available via web interfaces
and we note that the Mercury compiler, designed with industrial goals in mind,
includes a termination analysis (described in \cite{Speirs97a})
available as a compiler option.

On the other hand, non-termination analysis
seems to remain a much less attractive
subject. We can divide this line of research into two kinds of approaches:  dynamic
versus static analysis. In the former one, \cite{Bol91b} sets up some solid
foundations for loop checking, while \cite{Shen01a} presents some recent work.
The main idea is to \emph{prune} infinite derivations
\emph{at runtime} (some finite derivations may also be pruned by some
loop checkers).
In the latter approach, which
includes the work we present in this article, one tries to
\emph{compute at compile-time}
queries which admit at least one infinite derivation. 
One of the earliest works on the static approach is described
in~\cite{DeSchreye89a} 
where the authors
present an algorithm for detecting non-terminating atomic queries with
respect to (w.r.t.) a binary clause
of the form $p(\tilde{s})\leftarrow  p(\tilde{t})$. 
The condition is 
described  in terms of rational trees, while we aim at generalizing 
non-termination analysis for the generic CLP(X) framework.
Non-termination has also been studied in other paradigms, such as
Term Rewrite
Systems~\cite{Waldmann04,Giesl05,Zantema05,Waldmann07,Zankl07b,Payet08a};
the technique described in~\cite{Payet08a} is close to that
of this paper. In~\cite{Gupta08}, non-termination of C programs is considered
and in \cite{Godefroid05,Sen05}
some techniques are provided that detect crashes, assertion violation and
non-termination in C programs.

Our analysis shares with the work on termination analysis which is
presented in \cite{Codish99a}
a key component: the binary unfoldings of a logic program \cite{Gabbrielli94a},
which transform a finite set of definite clauses into a possibly infinite set of
facts and binary definite clauses.
Some termination analyses compute a finite over-approximation of the binary
unfolding semantics, over a constraint domain such as CLP($\cal{N}$).
In contrast, the non-termination analysis we have presented in~\cite{Payet06a}
starts from a finite subset $\mathit{BP}$ of the binary unfoldings of the
concrete program $P$; of course, a larger subset may increase the precision of
the analysis (\cite{Payet06a}~provides some experimental evidence).
This non-termination analysis first detects patterns of non-terminating
atomic queries from the binary recursive clauses and
then propagates this non-termination information to compute classes of
atomic queries for which we have a finite proof that there exists at least
one infinite derivation w.r.t. $\mathit{BP}$.
The equivalence between the termination of a logic program and that
of its binary unfoldings~\cite{Codish99a}
is a corner stone of the analysis; it allows us to conclude that any atomic
query belonging to the identified above classes admits an infinite left
derivation w.r.t. $P$. The basic idea in~\cite{Payet06a}
relies on checking, for each recursive clause in
$\mathit{BP}$, that the body is more general than the head; if this test
succeeds, we can conclude that the head is an atomic query which has
an infinite derivation w.r.t. $\mathit{BP}$. A key observation consists in
considering \emph{neutral argument positions} \ie{} argument positions of
the predicate symbols defined in $P$ that do not have any effect on
the derivation process when they are filled with a term that
satisfies a given condition. The subsumption test presented
in~\cite{Payet06a} only considers the arguments that are
in the non-neutral positions and checks that the arguments in the neutral
positions satisfy their associated condition. This extension of the
classical subsumption test considerably increases the power of the
approach in the sense that it allows one to compute more 
classes of non-terminating atomic queries.

The initial motivation in~\cite{Payet06a} was to complement termination
analysis with non-termination inside the logic programming paradigm 
in order to detect optimal termination conditions expressed in a language
describing classes of queries.
Although we obtained interesting experimental results, the overall approach
remains quite syntactic, with an \emph{ad hoc} flavor and tight links to some
basic logic programming machinery such as the unification algorithm. So
in the present paper our aim is to generalize the approach 
to the  constraint logic programming (CLP) setting and
the main contribution of this work consists in 
a strict generalization of the logical criterion
defined in \cite{Payet04b}.

The paper is organized as follows.
In Section~\ref{sect-preliminaries} we give some preliminary definitions
and in Section~\ref{section-loop-inference-with-constraints}
we recall in CLP terms the subsumption test to detect looping queries.
In Section~\ref{section-loop-filters} we introduce the 
neutral argument positions;
the operational definition we give (Section~\ref{section-DN-operational})
is useless in practice, hence
we propose a sufficient condition for neutrality, expressed as a logical
formula related to the constraint binary clause under consideration
(Section~\ref{section-logical-characterization}).
For some constraint domains, we show that the condition is 
also necessary (Section~\ref{section-DN-implies-DNlog}).
Depending on the constraint theory, the validity of such a condition can
be automatically decided. In Section~\ref{section-computing-DN-filters}, 
we describe an algorithm that uses the logical formula of
the sufficient condition to compute neutral argument positions.
Finally, in Section~\ref{section-impl} we describe our prototype
and we conclude in Section~\ref{section-conclusion}.
The detailed proofs of the results can be found in the
appendices at the end of the paper.

Notice that our approach consists in
computing a finite subset $\mathit{BP}$ of the binary unfoldings of the
program of interest and then in inferring non-terminating queries using
$\mathit{BP}$ only; hence,
\emph{we deliberately choose to restrict the analysis
  to binary CLP rules and atomic CLP queries as the result we obtain can
  be lifted to full CLP}.
\section{Preliminaries}\label{sect-preliminaries}
For any non-negative integer $n$, $[1,n]$ denotes the set
$\{1,\dots,n\}$. If $n=0$, then $[1,n]=\emptyset$.
We  recall  some basic  definitions about CLP, see~\cite{JMMS98}
for more details.
From now on, we fix an infinite countable set $\mathcal{V}$ of
\emph{variables} together with
a \emph{signature} $\Sigma$, \ie{} a pair $\langle F,\Pi\rangle$
where $F$ is a set of \emph{function symbols} and $\Pi$ is a set of
\emph{predicate symbols} with $F\cap\Pi=\emptyset$ and
$(F\cup\Pi)\cap\mathcal{V}=\emptyset$.
Every element of $F\cup\Pi$ has an
\emph{arity} which is the number of its arguments. We write
$f/n\in F$ (resp. $p/n\in \Pi$) to denote that $f$ (resp. $p$) is an
element of $F$ (resp. $\Pi$) whose arity is $n\geq 0$. A \emph{constant symbol}
is an element of $F$ whose arity is 0.

A \emph{term} is
a variable, a constant symbol or an object of the
form $f(t_1,\dots,t_n)$ where $f/n\in F$, $n\geq 1$ and $t_1,\dots,t_n$ are terms.
%
An \emph{atomic proposition} is an element $p/0$ of $\Pi$ or an object of
the form $p(t_1,\dots,t_n)$ where $p/n\in\Pi$, $n\geq 1$ and $t_1,\dots,t_n$ are
terms. A first-order \emph{formula} on $\Sigma$ is built from atomic propositions
in the usual way using the logical connectives $\land$, $\lor$, $\lnot$,
$\rightarrow$, $\leftrightarrow$ and the quantifiers $\exists$ and $\forall$.
If $\phi$ is a formula and $W:=\{X_1,\dots,X_n\}$ is
a set of variables, then $\exists_W \phi$ (resp. $\forall_W \phi$) denotes the
formula $\exists X_1\dots \exists X_n \phi$
(resp. $\forall X_1\dots \forall X_n \phi$).
We let $\exists \phi$ (resp. $\forall \phi$) denote
the existential (resp. universal) closure of $\phi$.

We fix a \emph{$\Sigma$-structure} $\mathcal{D}$, \ie{} a pair
$\langle D,[\cdot]\rangle$ which is an interpretation of the
symbols in $\Sigma$.
The set $D$ is called the \emph{domain} of $\mathcal{D}$ and $[\cdot]$
maps each $f/0\in F$ to an element of $D$ and
each $f/n\in F$ with $n\geq 1$ to a function $[f]: D^n \rightarrow D$; 
each $p/0\in \Pi$ to an element of $\{0,1\}$ and
  each $p/n\in\Pi$ with $n\geq 1$ to a boolean function
  $[p]: D^n \rightarrow \{0,1\}$.
We assume that the predicate symbol $=$ is in $\Sigma$ and 
is interpreted as identity in $D$.
%
A \emph{valuation} is a mapping from $\mathcal{V}$ to $D$.
Each valuation $v$ extends by morphism to terms.
As usual, a valuation $v$ induces a valuation $[\cdot]_v$ of
terms to $D$ and of formulas to $\{0,1\}$.

Given a formula $\phi$ and a valuation $v$, we write
$\mathcal{D}\models_v \phi$ when $[\phi]_v=1$.
We write $\mathcal{D}\models \phi$ when $\mathcal{D}\models_v \phi$ for all
valuation $v$. Notice that $\mathcal{D}\models \forall \phi$ if and
only if $\mathcal{D}\models \phi$, that $\mathcal{D}\models \exists \phi$ if
and only if there exists a valuation $v$ such that $\mathcal{D}\models_v \phi$,
and that $\mathcal{D}\models \lnot\exists \phi$ if and only if
$\mathcal{D}\models \lnot \phi$.
%
We say that a formula $\phi$ is \emph{satisfiable} (resp.
\emph{unsatisfiable}) in $\mathcal{D}$ when $\mathcal{D}\models \exists \phi$ (resp.
$\mathcal{D}\models \lnot \phi$).

We fix a set $\mathcal{L}$ of admitted formulas, the elements of which are
called \emph{constraints}.
We suppose that $\mathcal{L}$ is closed under variable renaming, existential
quantification and conjunction and that it contains all the atomic propositions,
the always satisfiable formula $\mathit{true}$ and the unsatisfiable formula
$\mathit{false}$.
We assume that there is a computable function $\mathit{solv}$ which maps each
$c\in\mathcal{L}$ to one of \texttt{true} or \texttt{false} indicating whether $c$
is satisfiable or unsatisfiable in $\mathcal{D}$.
We call $\mathit{solv}$ the \emph{constraint solver}.

\begin{example}[$\mathcal{Q}_{\mathit{lin}}$]
  \label{example-reals}
  The constraint domain $\mathcal{Q}_{\mathit{lin}}$
  has $<$, $\leq$, $=$, $\geq$, $>$ as predicate symbols,
  $+$, $-$, $*$, $/$  as function symbols and sequences of
  digits 
  as constant symbols.
  Only linear constraints are admitted. The domain of
  computation is the structure with the set of rationals,
  denoted by $\Q$, as domain and
  where the predicate symbols and the function symbols are
  interpreted as the usual relations and functions over the
  rationals. A constraint solver for $\mathcal{Q}_{\mathit{lin}}$
  always returning either {\tt true} or {\tt false} is described
  in~\cite{RVH96}. 
\end{example}


Sequences of distinct variables are denoted by
$\tilde{X}$, $\tilde{Y}$ or $\tilde{Z}$ and
are sometimes considered as sets of variables:
we may write $\forall_{\tilde{X}}$, $\exists_{\tilde{X}}$
or $\tilde{X}\cup\tilde{Y}$.
Sequences of (not necessarily distinct) terms are
denoted by $\tilde{s}$, $\tilde{t}$ or $\tilde{u}$. 
Given two sequences of $n$ terms
$\tilde{s}:=(s_1,\dots,s_n)$ and
$\tilde{t}:=(t_1,\dots,t_n)$, we write
$\tilde{s}=\tilde{t}$ either to denote the constraint
$s_1=t_1 \land \dots \land s_n=t_n$ or as a
shorthand for ``$s_1=t_1$ and \dots{} and $s_n=t_n$''.
Given a valuation $v$, we write
$v(\tilde{s})$ to denote the sequence
$(v(s_1),\dots,v(s_n))$ and
$[\tilde{s}]_v$ to denote the sequence
$([s_1]_v,\dots,[s_n]_v)$.

The signature in which all programs and queries under consideration are included
is $\Sigma_L := \langle F, \Pi\cup\Pi'\rangle$ where $\Pi'$ is the set of predicate
symbols that can be defined in programs, with $\Pi\cap\Pi'=\emptyset$. 

An \emph{atom} has the form $p(t_1,\dots,t_n)$ where $p/n\in\Pi'$ and
$t_1,\dots,t_n$ are terms. A \emph{program} is a finite set of clauses.
A \emph{clause} has the form $H \leftarrow c \diamond B$ where $H$ and $B$ are
atoms and $c$ is a finite conjunction of atomic propositions such
that $\mathcal{D}\models \exists c$. A \emph{query} has the form $\query{A}{d}$
where $A$ is an atom and $d$ is a finite conjunction of atomic
propositions.
Given an atom $A:=p(\tilde{t})$, we write $\rel{A}$ to denote the predicate
symbol $p$. Given a query $Q:=\query{A}{d}$, we write $\rel{Q}$ to denote the
predicate symbol $\rel{A}$. The set of variables occurring in some syntactic
objects $O_1,\dots,O_n$ is denoted $\Var(O_1,\dots,O_n)$.


We consider the following operational semantics given in
terms of \emph{derivations} from queries to queries.
Let $\query{p(\tilde{u})}{d}$ be a query
and $p(\tilde{s})\leftarrow c \diamond q(\tilde{t})$ be a 
fresh copy of a clause $r$.
When $\mathit{solv}(\tilde{s} = \tilde{u}
\land c \land d) = \mathtt{true}$ then
\[\query{p(\tilde{u})}{d} \lra_r
\query{q(\tilde{t})}{\tilde{s}=\tilde{u}\land c\land d}\]
is a  \emph{derivation step} of $\query{p(\tilde{u})}{d}$
w.r.t. $r$ with $p(\tilde{s})\leftarrow c \diamond q(\tilde{t})$ as its
\emph{input clause}. We write $Q \lra_P^+ Q'$ to summarize a finite number
($> 0$) of derivation steps from $Q$ to $Q'$
where each input clause is a variant of a clause from
program $P$. 
Let $Q_0$ be a query. A sequence of derivation steps
$Q_0 \lra_{r_1} Q_1 \lra_{r_2} \cdots$ of maximal
length is called a \emph{derivation}
of $P\cup\{Q_0\}$ when $r_1$, $r_2$, \dots are clauses from $P$
and the \emph{standardization apart} condition holds, \ie{}
each input clause used is variable disjoint from the
initial query $Q_0$ and from the input clauses
used at earlier steps.
We say $Q_0$ \emph{loops} w.r.t. $P$
when there exists an infinite derivation of $P\cup\{Q_0\}$.


\section{Loop Inference with Constraints}
\label{section-loop-inference-with-constraints}
In the logic programming framework, the subsumption test provides
a simple way to infer looping queries: if, in a logic
program $P$, there is a clause $p(\tilde{s})\leftarrow p(\tilde{t})$
such that $p(\tilde{t})$ is more general than $p(\tilde{s})$,
then the query $p(\tilde{s})$ loops w.r.t. $P$.
In this section, we extend this result to the constraint logic
programming framework.

\subsection{A ``More General Than'' Relation}
\label{section-set-query}

A query can be viewed as a finite description of a possibly
infinite set of atoms, the arguments of which are values
from $D$.
%
\begin{example}
  In the constraint domain $\mathcal{Q}_{\mathit{lin}}$,
  the query $Q:=\query{p(X,Y)}{Y\leq X+2}$ describes
  the set of atoms $p(x,y)$ where $x$ and $y$ are rational numbers
  and $X$ and $Y$ can be made equal to $x$ and $y$ respectively
  while the constraint $Y\leq X+2$ is satisfied. For instance,
  $p(0,2)$ is an element of the set described by $Q$.
\end{example}
In order to capture this intuition, we introduce the
following definition.
%
\begin{definition}[Set Described by a Query]
  The set of atoms that is described by a query
  $Q:=\query{p(\tilde{t})}{d}$ is denoted by
  $\Set{Q}$ and is defined as:
  $\Set{Q} = \{p([\tilde{t}]_v) \; | \; \mathcal{D} \models_v d\}$.
\end{definition}

Clearly, $\Set{\query{p(\tilde{t})}{d}}=\emptyset$ if and
only if $d$ is unsatisfiable in $\mathcal{D}$.
Moreover, two variants describe the same set:
%
\begin{lemma}\label{lemma-set-variant}
  Let $Q$ and $Q'$ be two queries such that
  $Q'$ is a variant of $Q$. Then, $\Set{Q}=\Set{Q'}$.
\end{lemma}

Notice that the operational semantics we introduced above
can be expressed using sets described by queries:
%
\begin{lemma}\label{lemma-operational-sem}
  Let $Q$ be a query and $r:=H\leftarrow c\diamond B$ be a clause.
  There exists a derivation step of $Q$ w.r.t. $r$
  if and only if  $\Set{Q}\cap\Set{\query{H}{c}}\neq\emptyset$.
\end{lemma}

The ``more general than'' relation we consider is defined as follows:
%
\begin{definition}[More General]
  We say that a query $Q_1$ is \emph{more general than}
  a query $Q$ when $\Set{Q}\subseteq \Set{Q_1}$.
\end{definition}
%
\begin{example}
  In $\mathcal{Q}_{\mathit{lin}}$, the query 
  $Q_1:=\query{p(X,Y)}{Y\leq X+3}$ is more general than
  the query $Q:=\query{p(X,Y)}{Y\leq X+2}$. However,
  $Q$ is not more general than $Q_1$; for instance,
  $p(0,3)\in\Set{Q_1}$ but $p(0,3)\not\in\Set{Q}$.
\end{example}

\subsection{Loop Inference}
\label{section-loop-constraints}

Suppose we have a derivation step $Q\lra_r Q_1$ where
$r:=H\leftarrow c\diamond B$. Then,
by Lemma~\ref{lemma-operational-sem},
$\Set{Q}\cap\Set{\query{H}{c}}\neq\emptyset$.
Hence, if $Q'$ is a query that is more general than
$Q$, as $\Set{Q}\subseteq\Set{Q'}$, we have
$\Set{Q'}\cap\Set{\query{H}{c}}\neq\emptyset$. So,
by Lemma~\ref{lemma-operational-sem}, there exists
a query $Q'_1$ such that $Q'\lra_r Q'_1$.
The following lifting result says that, moreover, $Q'_1$ is
more general than $Q_1$.
%
\begin{theorem}[Lifting]\label{theorem-lifting}
  Consider a derivation step $Q \lra_r Q_1$ and
  a query $Q'$ that is more general than $Q$.
  Then, there exists a derivation step $Q' \lra_r Q'_1$
  where $Q'_1$ is more general than $Q_1$.
\end{theorem}

From this theorem, we derive two corollaries that
can be used to infer looping queries just from the
text of a program.
%
\begin{corollary}
  \label{coro-p-if-p}
  Let $r:=H\leftarrow c\diamond B$ be a clause.
  If $\query{B}{c}$ is more general than $\query{H}{c}$ then
  $\query{H}{c}$ loops w.r.t. $\{r\}$.
\end{corollary}
The intuition of Corollary~\ref{coro-p-if-p} is that we have
$\query{H}{c}\lra_r Q_1$ where $Q_1$ is a variant of $\query{B}{c}$;
hence, $Q_1$ is more general than $\query{H}{c}$;
so, by the Lifting Theorem~\ref{theorem-lifting},
there exists a derivation step $Q_1\lra_r Q_2$ where
$Q_2$ is more general than $Q_1$; by repeatedly using this
reasonning, one can build an infinite derivation of
$\{r\}\cup\{\query{H}{c}\}$.
%
\begin{corollary}
  \label{coro-p-if-q}
  Let $r:=H\leftarrow c\diamond B$
  be a clause from a program $P$.
  If $\query{B}{c}$ loops w.r.t.
  $P$ then $\query{H}{c}$ loops w.r.t. $P$.
\end{corollary}
The intuition of Corollary~\ref{coro-p-if-q} is that we have
$\query{H}{c}\lra_r Q_1$ where $Q_1$ is a variant of $\query{B}{c}$,
which implies that $Q_1$ is more general than $\query{B}{c}$; as
there exists an infinite derivation $\xi$ of $P\cup\{\query{B}{c}\}$,
by successively applying the Lifting
Theorem~\ref{theorem-lifting}
to each step of $\xi$ one can construct an infinite
derivation of $P\cup\{Q_1\}$.
%
\begin{example}
  Consider the following recursive clause $r$ in $\mathcal{Q}_{\mathit{lin}}$:
  \[p(N) \la N \geq 1 \land N = N_1+1 \; \diamond p(N_1)\]
  The query
  $Q_1:=\query{p(N_1)}{N \geq 1 \land N = N_1+1 }$
  is more general than the query
  $Q:=\query{p(N)}{N \geq 1 \land N = N_1+1}$
  (for instance, $p(0)\in\Set{Q_1}$ but $p(0)\not\in\Set{Q}$).
  So, by Corollary~\ref{coro-p-if-p}, $Q$ loops w.r.t.
  $\{r\}$. Therefore, there exists an infinite derivation $\xi$ of
  $\{r\}\cup\{Q\}$. Then, if $Q'$ is a query that is more general than
  $Q$, by successively applying the Lifting Theorem~\ref{theorem-lifting}
  to each step of $\xi$, one can construct an infinite derivation of
  $\{r\}\cup\{Q'\}$. So, $Q'$ also loops w.r.t. $\{r\}$. 
\end{example}


\section{Loop Inference Using Filters}
\label{section-loop-filters}

The condition provided by
Corollary~\ref{coro-p-if-p} is rather weak because it fails
at inferring looping queries in some simple cases.
This is illustrated by the following example.
%
\begin{example}
  \label{example-neutral-Rlin}
  Consider the following recursive clause $r$ in $\mathcal{Q}_{\mathit{lin}}$:
  \[p(N,T) \la N \geq 1 \land N = N_1+1 \land T_1=2*T \land T \geq 1
  \; \diamond p(N_1,T_1)\]
  Let $c$ denote the constraint in $r$.
  The query $\query{p(N,T)}{c}$ loops w.r.t.
  $\{r\}$ because only the first argument of $p$
  decreases in $r$ and in this query it is unspecified.
  But we cannot infer that $\query{p(N,T)}{c}$ loops
  w.r.t. $\{r\}$ from Corollary~\ref{coro-p-if-p} 
  as in $r$ $\query{p(N_1,T_1)}{c}$ is not more
  general than $\query{p(N,T)}{c}$ because of the second
  argument of $p$: for instance, $p(1,1)\in\Set{\query{p(N,T)}{c}}$
  but $p(1,1)\not\in\Set{\query{p(N_1,T_1)}{c}}$.
\end{example}

In what follows, we extend the relation ``is more general''.
Instead of comparing atoms in all positions using the ``more general''
relation, we distinguish some predicate argument positions
for which we just require that a certain property must hold, while
for the other positions we use the ``more general'' relation as
before. Doing so, we aim at inferring more looping queries.
%
\begin{example}[Example~\ref{example-neutral-Rlin} continued]
  \label{example-neutral-Rlin-continued}
  Let us consider argument position $2$ of predicate symbol $p$.
  In the clause $r$, the projection of $c$ on $T$
  is equivalent to $T\geq 1$; this projection expresses the constraint
  placed upon the second argument of $p$ to get a derivation step
  with $r$. Notice that the projection of $c$ on $T_1$ is equivalent
  to $T_1\geq 2$, which implies $T_1\geq 1$. Therefore, the requirements
  on the head variable $T$ propagates to the body variable $T_1$.
  Moreover, the ``piece'' $\query{p(N_1)}{c}$ of $\query{p(N_1,T_1)}{c}$
  is more general than the ``piece'' $\query{p(N)}{c}$ of
  $\query{p(N,T)}{c}$. Consequently, $\query{p(N_1,T_1)}{c}$ is
  more general than $\query{p(N,T)}{c}$ up to the second argument of $p$
  which, in $\query{p(N_1,T_1)}{c}$, satisfies $T_1\geq 1$, the condition
  to get a derivation step with $r$. Hence, by an extended version of
  Corollary~\ref{coro-p-if-p} we could infer that $\query{p(N,T)}{c}$
  loops w.r.t. $\{r\}$.
\end{example}

\subsection{Sets of Positions}
\label{section-sets-of-pos}
A basic idea in Example~\ref{example-neutral-Rlin-continued}
lies in identifying argument positions of predicate symbols.
Below, we introduce a formalism to do so.

\begin{definition}[Set of Positions]
  A \emph{set of positions}, denoted by $\tau$,
  is a function that maps each $p/n \in \Pi'$ to a subset of
  $[1,n]$.
\end{definition}

\begin{example}
  \label{example-set-of-pos}
  If we want to distinguish the second argument position of
  the predicate symbol $p$ defined in
  Example~\ref{example-neutral-Rlin}, we set
  $\tau := \langle p \mapsto\{2\}\rangle$.
  If we do not want to distinguish any argument position of
  $p$, we set $\tau' := \langle p \mapsto\emptyset\rangle$.
\end{example}

\begin{definition}
  Let $\tau$ be a set of positions. Then, $\overline{\tau}$
  is the set of positions defined as: for each $p/n \in \Pi'$,
  $\overline{\tau}(p) = [1,n]\setminus\tau(p)$.
\end{definition}

\begin{example}
  If we set
  $\tau := \langle p \mapsto\{2\}\rangle$ and
  $\tau' := \langle p \mapsto\emptyset\rangle$
  where the arity of $p$ is 2,
  then
  $\overline{\tau} = \langle p\mapsto\{1\}\rangle$
  and
  $\overline{\tau}' = \langle p\mapsto\{1,2\}\rangle$.
\end{example}

Using a set of positions $\tau$, one can \emph{project}
syntactic objects: 
\begin{definition}[Projection] \label{def-projection}
  Let $\tau$ be a set of positions.
  \begin{itemize}
  \item The projection of $p\in\Pi'$ on $\tau$ is the
    predicate symbol denoted by $\restrict{p}{\tau}$.
    Its arity is the number of elements of $\tau(p)$.
  \item Let $p/n\in \Pi'$
    and $\tilde{t}:=(t_1,\dots,t_n)$ be a sequence of
    $n$ terms.
    The \emph{projection of $\tilde{t}$ on $\tau(p)$},
    denoted by $\restrict{\tilde{t}}{\tau(p)}$, is the
    sequence $(t_{i_1},\dots,t_{i_m})$ where
    $\{i_1,\dots,i_m\}=\tau(p)$ and
    $i_1<\dots<i_m$.
  \item Let $A:=p(\tilde{t})$ be an atom. The projection
    of $A$ on $\tau$, denoted by $\restrict{A}{\tau}$, is the
    atom $\restrict{p}{\tau}(\restrict{\tilde{t}}{\tau(p)})$.
  \item The projection of a query $\query{A}{d}$ on $\tau$,
    denoted by $\restrict{\query{A}{d}}{\tau}$,
    is the query $\query{\restrict{A}{\tau}}{d}$.
  \end{itemize}
\end{definition}

\begin{example}[Example~\ref{example-set-of-pos} continued]
  \label{example-restriction}
  The projection of the query $\query{p(N,T)}{c}$ on $\tau$
  (resp. $\tau'$) is the query $\query{p_{\tau}(T)}{c}$
  (resp. the query $\query{p_{\tau'}}{c}$).
\end{example}

Projection preserves inclusion and non-disjointness of sets
described by queries:
\begin{lemma}[Inclusion]
  \label{lemma-restriction}
  Let $\tau$ be a set of positions and $Q$ and $Q'$ be
  two queries. If $\Set{Q}\subseteq\Set{Q'}$ then
  $\Set{\restrict{Q}{\tau}}\subseteq\Set{\restrict{Q'}{\tau}}$.
\end{lemma}
%
\begin{lemma}[Non-Disjointness]
  \label{lemma-useful2-theo-DNlog-iff-DN}
  Let $\tau$ be a set of positions and
  $Q$ and $Q'$ be two queries.
  If $\Set{Q}\cap\Set{Q'}\neq\emptyset$
  then $\Set{\restrict{Q}{\tau}}\cap
  \Set{\restrict{Q'}{\tau}}\neq\emptyset$.
\end{lemma}
%

\subsection{Filters}
A second idea in Example~\ref{example-neutral-Rlin-continued} consists
in associating constraints with argument positions ($T\geq 1$ for
position 2 in Example~\ref{example-neutral-Rlin-continued}). We define
a filter to be the combination of sets of positions with their
associated constraint:
\begin{definition}[Filter]\label{def-filter}
  A \emph{filter}, denoted by $\Delta$, is a pair $(\tau,\delta)$
  where $\tau$ is a set of positions and $\delta$ is a
  function that maps each $p \in \Pi'$ to a query of the form
  $\query{\restrict{p}{\tau}(\tilde{t})}{d}$ where
  $\mathcal{D} \models \exists d$.
\end{definition}

\begin{example}
  \label{ex-filter-Rlin}
  Consider $\tau := \langle p \mapsto\{2\}\rangle$ and
  $\tau' := \langle p \mapsto\emptyset\rangle$.
  Let $\delta := \langle \; p \mapsto
  \query{\restrict{p}{\tau}(B)}
  {B\geq 1} \; \rangle$ and
  $\delta' := \langle \; p \mapsto
  \query{\restrict{p}{\tau'}}{\mathit{true}} \; \rangle$.
  Then, $\Delta:=(\tau,\delta)$ and $\Delta':=(\tau',\delta')$
  are filters.
\end{example}
Note that $\delta(p)$ is given in the form of a query
$\query{\restrict{p}{\tau}(\tilde{t})}{d}$,
instead of just a constraint $d$, because we need to indicate that
the entry points of $d$ are the terms in $\tilde{t}$. Indeed,
the function $\delta$ is used to ``filter'' queries: we say that
a query $Q$ \emph{satisfies} $\Delta$  when the set of atoms
described by $\restrict{Q}{\tau}$, the projection of $Q$ on the positions
$\tau$, is included in the set of atoms described by
$\delta(\rel{Q})$, the query defined for $Q$'s predicate symbol by $\Delta$.
More formally:
\begin{definition}[Satisfies]
  \label{def-satisfies}
  Let $\Delta:=(\tau,\delta)$ be a filter and
  $Q$ be a query. Let $p:=\mathit{rel}(Q)$.
  We say that $Q$ \emph{satisfies} $\Delta$ when
  $\Set{\restrict{Q}{\tau}}\subseteq\Set{\delta(p)}$.
\end{definition}

Now we come to the extension of the relation
``more general than''. 
Intuitively, $\query{p(\tilde{t'})}{d'}$ is
$\Delta$-more general than $\query{p(\tilde{t})}{d}$ if the ``more
general than'' relation holds for the elements of $\tilde{t}$
and $\tilde{t'}$ whose position is not in $\tau$ while
the elements of $\tilde{t'}$ whose position is in $\tau$ satisfy
$\delta$. More formally:
\begin{definition}[$\Delta$-More General]
  \label{def-Delta-more-gen-state}
  Let $\Delta:=(\tau,\delta)$ be a filter and
  $Q$ and $Q'$ be two queries.
  We say that $Q'$ is \emph{$\Delta$-more general than} $Q$
  when $\restrict{Q'}{\overline{\tau}}$ is more general than
  $\restrict{Q}{\overline{\tau}}$ and $Q'$ satisfies
  $\Delta$.
\end{definition}
%
\begin{example}
  \label{ex-delta-more-general}
  Consider the constraint $c$ in the clause
  \[\mathit{p}(N,T) \la N\geq 1\land N = N_1+1 \land T_1=2*T \land T\geq 1
  \diamond p(N_1,T_1)\]
  of Example~\ref{example-neutral-Rlin}.
  The query $Q_1:=\query{p(N_1,T_1)}{c}$
  is $\Delta$-more general than
  $Q:=\query{p(N,T)}{c}$
  for the filter
  $\Delta:=\big(\big\langle p \mapsto\{2\}\big\rangle,
  \big\langle p \mapsto
  \query{\restrict{p}{\tau}(B)}
  {B\geq 1} \big\rangle\big)$. 
  However, $Q_1$ is not
  $\Delta'$-more general than $Q$ for the filter
  $\Delta':=\big(\big\langle p \mapsto\emptyset\big\rangle,
  \big\langle p \mapsto
  \query{\restrict{p}{\tau'}}
  {\mathit{true}} \big\rangle\big)$;
  indeed,
  $\tau'(p)=\emptyset$ implies that being
  $\Delta'$-more general is equivalent to being more
  general and, by Example~\ref{example-neutral-Rlin},
  $Q_1$ is not more general than $Q$.
\end{example}

\begin{lemma}[Transitivity]
  \label{lemma-properties-delta}
  For any filter $\Delta$, the ``$\Delta$-more general than'' relation
  is transitive.
\end{lemma}

Notice that for any filter $\Delta:=(\tau,\delta)$ and any query $Q$,
we have that $\restrict{Q}{\overline{\tau}}$ is more general than itself
(because the ``more general than'' relation is reflexive), but $Q$ may
not satisfy $\Delta$. Hence, the ``$\Delta$-more general than'' relation
is not always reflexive.
\begin{example}
  Consider the constraint domain $\mathcal{Q}_{\mathit{lin}}$. Let
  $p/1 \in\Pi'$ and $\Delta:=(\tau,\delta)$ be the filter defined by
  $\tau:= \langle p \mapsto \{1\}\rangle$ and
  $\delta := \langle \; p \mapsto \query{\restrict{p}{\tau}(X)}
  {X\geq 1} \; \rangle$.
  The query $Q:=\query{p(0)}{\mathit{true}}$ is not $\Delta$-more general
  than itself because
  $\Set{\restrict{Q}{\tau}}=\{\restrict{p}{\tau}(0)\}
  \not\subseteq\{\restrict{p}{\tau}(x) \;|\; x \text{ is a rational and }
  x\geq 1\}=\Set{\delta(p)}$.
  Hence, $Q$ does not satisfy $\Delta$. 
\end{example}

The fact that reflexivity does not always hold is
an expected property. Indeed,
suppose that a filter $\Delta:=(\tau,\delta)$ induces
a ``$\Delta$-more general than'' relation  that is
reflexive. Then for any queries $Q$ and $Q'$,
we have that $Q'$ is $\Delta$-more general
than $Q$ if and only if $\restrict{Q'}{\overline{\tau}}$ is
more general than $\restrict{Q}{\overline{\tau}}$
(because, as $Q'$ is $\Delta$-more general than itself,
$Q'$ necessarily satisfies $\Delta$). Hence, $\delta$ is useless in the
sense that it ``does not filter anything''.
Filters equipped with such a $\delta$ 
were introduced in~\cite{Payet04b} where for any
predicate symbol $p$, $\delta(p)$ has the form
$\query{\restrict{p}{\tau}(\tilde{X})}{\mathit{true}}$,
where $\tilde{X}$ is a sequence of distinct variables.
In this paper, we aim at generalizing the approach
of~\cite{Payet04b}. Hence, we also consider functions
$\delta$ that really filter queries.

\subsection{DN Filters: an Operational Definition}
\label{section-DN-operational}
Let us now introduce a special kind of filters that we
call ``derivation neutral''.
The name ``derivation neutral'' stems from the fact
that if in a derivation of a query $Q$, we replace
$Q$ by a $\Delta$-more general $Q'$,
then we get a ``similar'' derivation.
%
\begin{definition}[Derivation Neutral]
  \label{def-DN-filter}
  Let $r$ be a clause and $\Delta$ be a filter.
  We say that $\Delta$ is \emph{DN}
  for $r$ when for each derivation step $Q \lra_r Q_1$,
  the query $Q_1$ satisfies $\Delta$ and
  for each query $Q'$ that is $\Delta$-more general than $Q$,
  there exists a derivation step $Q'\lra_r Q'_1$ 
  where $Q'_1$ is $\Delta$-more general than $Q_1$.
  This definition is extended to programs:
  $\Delta$ is \emph{DN} for $P$
  when it is DN for each clause of $P$.
\end{definition}

Derivation neutral filters lead to the following extended version
of Corollary~\ref{coro-p-if-p} (to get Corollary~\ref{coro-p-if-p},
take $\Delta:=(\tau,\delta)$ with $\tau(p)=\emptyset$ for any $p$).
%
\begin{theorem}
  \label{propo-p-if-p-Delta}
  Let $r := H \leftarrow c \diamond B$ be a clause. 
  Let $\Delta$ be a filter that is DN for $r$.
  If $\query{B}{c}$ is $\Delta$-more
  general than $\query{H}{c}$ then
  $\query{H}{c}$ loops w.r.t. $\{r\}$.
\end{theorem}
%
\begin{example}
  If the filter
  $\Delta$
  of Example~\ref{ex-delta-more-general}
  is DN for the clause $r=p(N,T)\leftarrow c\diamond p(N_1,T_1)$
  of Example~\ref{example-neutral-Rlin}, then
  we can deduce that $\query{p(N,T)}{c}$
  loops w.r.t. $\{r\}$ because $\query{p(N_1,T_1)}{c}$
  is $\Delta$-more general than $\query{p(N,T)}{c}$ (see
  Example~\ref{ex-delta-more-general}). 
\end{example}

Computing a derivation neutral filter from the text of a program
is not straightforward if we use the above definition. 
Section~\ref{section-logical-characterization} presents a logical
characterization that we use in Section~\ref{section-computing-DN-filters} 
to compute a filter that is DN for a given recursive clause.

\subsection{A Logical Characterization of DN Filters}
\label{section-logical-characterization}
From now on, we suppose, without loss
of generality, that a clause has the form 
$p(\tilde{X}) \leftarrow c \diamond q(\tilde{Y})$ where
$\tilde{X}$ and $\tilde{Y}$ are disjoint sequences of
distinct variables. Hence, $c$ is the conjunction of
all the constraints, including unifications.
We distinguish the following set of variables
that appear inside such a clause.
%
\begin{definition}
  The set of \emph{local variables} of a clause
  $r:=p(\tilde{X}) \leftarrow c \diamond q(\tilde{Y})$
  is $\mathit{local\_vars}(r) := \Var(c) \setminus (
  \tilde{X} \cup \tilde{Y})$.
\end{definition}

In this section, we aim at characterizing DN filters in a logical
way. To this end, we define:
%
\begin{definition}[sat]
  Let $Q:=\query{p(\tilde{t})}{d}$ be a query and
  $\tilde{s}$ be a sequence of terms of the same length as $\tilde{t}$.
  Then, $\sat{\tilde{s}}{Q}$ denotes a  formula of the form
  $\exists_{\Var(Q')} (\tilde{s}=\tilde{t}'\land d')$
  where $Q':=\query{p(\tilde{t}')}{d'}$ is a variant of $Q$
  and variable disjoint with $\tilde{s}$.
\end{definition}
Intuitively, $\sat{\tilde{s}}{Q}$ holds when the terms in the sequence
$\tilde{s}$ satisfy the constraint $d$, the entry points of which are
the terms in $\tilde{t}$.
Clearly, the satisfiability of $\sat{\tilde{s}}{Q}$ does not depend
on the choice of the variant of $Q$.
The set that is described by a query can then be
characterized as follows:
\begin{lemma}\label{lemma-sat-set}
  Let $Q$ be a query and $p:=\rel{Q}$. Let $\tilde{u}$ be
  a sequence of $\arity{p}$ terms and $v$ be a
  valuation. Then, $p([\tilde{u}]_v)\in\Set{Q}$ if and only if
  $\mathcal{D} \models_v \sat{\tilde{u}}{Q}$.
\end{lemma}

Now we give a logical definition of derivation neutrality.
As we will see later, under certain circumstances, this
definition is equivalent to the operational one we
gave above.
\begin{definition}[Logical Derivation Neutral]
  \label{def-log-DN}
  We say that a filter $\Delta:=(\tau,\delta)$ is
  \emph{DNlog} for a clause
  $r:=p(\tilde{X}) \leftarrow c \diamond q(\tilde{Y})$ when
  \[\mathcal{D} \models
  c \rightarrow \forall_{\restrict{\tilde{X}}{\tau(p)}} \big[
  \sat{\restrict{\tilde{X}}{\tau(p)}}{\delta(p)} \rightarrow
  \exists_{\mathcal{Y}} c
  \big]
  \quad
  \text{and}
  \quad
  \mathcal{D}\models c\rightarrow
  \sat{\restrict{\tilde{Y}}{\tau(q)}}{\delta(q)}\]
  where $\mathcal{Y} = \restrict{\tilde{Y}}{\tau(q)}
  \cup \mathit{local\_vars}(r)$.
\end{definition}
%
\begin{example}
  \label{example-DNlog}
  In $\mathcal{Q}_{\mathit{lin}}$, the filter
  $\big(\big\langle p \mapsto\{2\}\big\rangle,
  \big\langle p \mapsto
  \query{\restrict{p}{\tau}(B)}
  {B\geq 1} \big\rangle\big)$
  is DNlog for the clause
  \[\mathit{p}(N,T) \la N\geq 1\land N = N_1+1 \land T_1=2*T \land T\geq 1
  \diamond p(N_1,T_1)\] of
  Example~\ref{example-neutral-Rlin}.
  Indeed, $\restrict{\tilde{X}}{\tau(p)}=\{T\}$,
  $\restrict{\tilde{Y}}{\tau(q)}=\{T_1\}$ and
  $\mathit{local\_vars}(r)=\{\}$. So, if we let $c$
  denote the constraint in this clause,
  the formulas of Definition~\ref{def-log-DN} turn into
  \[
  \mathcal{D} \models
  c \rightarrow \forall T\,
  \big[T\geq 1 \rightarrow
  \exists T_1\,  c\big]
  \quad
  \text{and}
  \quad
  \mathcal{D}\models c\rightarrow
  T_1\geq 1\] which are true.
\end{example}

The first formula in Definition~\ref{def-log-DN} has the
following meaning. If one holds a solution for constraint $c$,
then, changing the value given to the variables of $\tilde{X}$
distinguished by $\tau$ to some value satisfying $\delta(p)$,
there exists a value for the local variables and the
variables of $\tilde{Y}$ distinguished by $\tau$ such that
$c$ is still satisfied. This formula expresses
the fact that DNlog arguments (\ie{} those distinguished by
$\tau$) \emph{do not interact} in $c$ with the other arguments.
Intuitively, two variables $X_1$ and $X_2$ do not interact in
a constraint $c$ when the set of values assigned to $(X_1,X_2)$ by
all the solutions of $c$
results from the exhaustive combination of the set of values
assigned to $X_1$ by all the solutions of $c$
and the set of values assigned to $X_2$
by all the solutions of $c$; more formaly, when
\[\big\{(v(X_1),v(X_2))\;|\;\mathcal{D}\models_v c\big\} = 
\big\{v(X_1)\;|\;\mathcal{D}\models_v c\big\}
\times
\big\{v(X_2)\;|\;\mathcal{D}\models_v c\big\}\;.
\]
%
%
\begin{example}
  \begin{itemize}
  \item In Example~\ref{example-DNlog} above, the set of values
    assigned to $(N,T)$ by all the solutions of $c$ is 
    $\{(a,b)\;|\; a\geq 1,\ b\geq 1\}$. We have
    $\{(a,b)\;|\; a\geq 1,\ b\geq 1\} =
    \{a\;|\; a\geq 1\}\times\{b\;|\; \ b\geq 1\}$ where
    $\{a\;|\; a\geq 1\}$ is the set of values
    assigned to $N$ by all the solutions of $c$ and
    $\{b\;|\; b\geq 1\}$ is the set of values
    assigned to $T$ by all the solutions of $c$. Hence,
    $N$ and $T$ do not interact.
  \item Now consider $c=(X\geq Z\land Z\geq Y)$.
    The set of values assigned to
    $(X,Y)$ by all the solutions of $c$ is 
    $\{(a,b)\;|\; a\geq b\}$ and 
    the set of values assigned to $X$ and to $Y$ by all the
    solutions of $c$ is $\Q$. As 
    $\{(a,b)\;|\; a\geq b\}\neq\Q\times\Q$, we have
    that $X$ and $Y$ do interact.
  \end{itemize}
\end{example}

%
The second formula in Definition~\ref{def-log-DN} means that any
solution of $c$ assigns to the variables of $\tilde{Y}$ distinguished
by $\tau$ a value that satisfies $\delta(q)$.
This corresponds to the intuition that neutral argument positions are
sorts of ``pipes'' where one can place any term satisfying $\delta$
with no effect on the derivation process.

The logical definition of derivation neutrality implies the
operational one:
%
\begin{theorem}
  \label{prop-DNlog-implies-DN}
  Let $r$ be a clause and $\Delta$ be a filter.
  If $\Delta$ is DNlog for $r$ then
  $\Delta$ is DN for $r$.
\end{theorem}
%

DNlog in Definition~\ref{def-log-DN} consists of two 
formulas, say DNlog1 and DNlog2, where
DNlog2 requires $\restrict{\tilde{Y}}{\tau(q)}$
to always satisfy $\delta(q)$. One may think of a (perhaps more natural)
requirement, say DNlog12, resulting from ``merging'' 
DNlog1 and DNlog2:
\[\mathcal{D} \models
c \rightarrow \forall_{\restrict{\tilde{X}}{\tau(p)}} \big[
\sat{\restrict{\tilde{X}}{\tau(p)}}{\delta(p)} \rightarrow
\exists_{\mathcal{Y}} (c \land \sat{\restrict{\tilde{Y}}{\tau(q)}}{\delta(q)})
\big]\;.
\]
The point is that a filter satisfying DNlog12 is not necessarily
DN (\ie{} Theorem~\ref{prop-DNlog-implies-DN} does not hold for 
DNlog12). For instance, consider in $\mathcal{Q}_{\mathit{lin}}$
the clause
\[r:=p(X)\leftarrow X\leq 3\land 2\leq Y \diamond p(Y)\]
and the filter 
$\Delta:=(\tau,\delta)$ with $\tau(p)=\{1\}$ and 
$\delta(p)=\query{\restrict{p}{\tau}(X)}{X\leq 3}$.
Then, DNlog2 \ie
\[\mathcal{D}\models c\rightarrow
\sat{\restrict{\tilde{Y}}{\tau(q)}}{\delta(q)}\]
does not hold: we have
$\restrict{\tilde{Y}}{\tau(p)}=\{Y\}$ 
and any valuation
$v$ with $v(X)=1$ and $v(Y)=4$ is a solution of the constraint $c$
in $r$ \ie{}
$\mathcal{D}\models_v c$;
but, as $3 < v(Y)$, we have
$\mathcal{D}\not\models_v Y\leq 3$ \ie{}
$\mathcal{D}\not\models_v \sat{\restrict{\tilde{Y}}{\tau(p)}}{\delta(p)}$;
therefore, $\mathcal{D}\not\models_v c\rightarrow
\sat{\restrict{\tilde{Y}}{\tau(p)}}{\delta(p)}$. Hence,
$\Delta$ is not DNlog for $r$.
In the next section (see Theorem~\ref{prop-DN-implies-DNlog2} and
Example~\ref{example-Qlin-DN-implies-DNlog}) we prove that DNlog
in $\mathcal{Q}_{\mathit{lin}}$ is equivalent to DN. Therefore, $\Delta$
is not DN for $r$. On the other hand, DNlog12 holds
as in this example it is equivalent to
(we have $\restrict{\tilde{X}}{\tau(p)}=\{X\}$ and
$\mathcal{Y}=\{Y\}$):
\[\mathcal{D}\models c\rightarrow
\forall_X\big[X\leq 3\rightarrow \exists_Y(c\land Y\leq 3)\big]\;.\]

\subsection{When DN Filters Are Also DNlog}
\label{section-DN-implies-DNlog}
DN filters are not always DNlog as illustrated by the
following example.
%
\begin{example}\label{example-DN-implies-DNlog}
  Suppose that $\Sigma=\{0,=,\geq\}$ and 
  $\mathcal{D}=\mathcal{D}_{\mathcal{Q}_{\mathit{lin}}}$.
  Consider
  \[r:=p(X_1,X_2)\leftarrow X_2\geq X_1\land X_1\geq 0 \land Y_1=X_1
  \land Y_2=X_2 \diamond p(Y_1,Y_2)\;.\]
  Let $c$ denote the constraint in $r$.
  Consider also a filter $\Delta:=(\tau,\delta)$ where $\tau(p)=\{1\}$ and
  $\delta(p)=\query{\restrict{p}{\tau}(X)}{X\geq 0}$.
  Notice that given the form of $\Sigma$, 
  one cannot write a constraint that has only one solution different
  from $0$; more precisely, for any terms $t_1$
  and $t_2$ and any constraint $d\neq\mathit{false}$:
  \begin{equation}\label{eqn:exDNnotDNlog}
    p(0,0) \in \Set{\query{p(t_1,t_2)}{d}}\;.
  \end{equation}
  %
  Whatever $Q$, if there is a derivation step $Q\lra_r Q_1$:
  \begin{itemize}
  \item the query $Q_1$ satisfies $\Delta$ because $c$ implies that
    $Y_1\geq 0$,
  \item for any  $Q'$ that is $\Delta$-more general than $Q$,
    $\Set{\query{p(X_1,X_2)}{c}}\cap \Set{Q'} \neq \emptyset$
    because by~(\ref{eqn:exDNnotDNlog})
    $p(0,0) \in \Set{\query{p(X_1,X_2)}{c}}\cap \Set{Q'}$;
    hence, there exists a derivation step $Q'\lra_r Q'_1$.
    Notice that $\restrict{{Q'_1}}{\overline{\tau}}$ is more general than
    $\restrict{{Q_1}}{\overline{\tau}}$ because 
    $\restrict{Q'}{\overline{\tau}}$ is more general than
    $\restrict{Q}{\overline{\tau}}$ and $c$ demands that
    $Y_2=X_2$; moreover, $Q'_1$ satisfies $\Delta$ because $c$ implies that
    $Y_1\geq 0$; therefore, $Q'_1$ is $\Delta$-more general than $Q_1$.
  \end{itemize}
  Consequently, $\Delta$ is DN for $r$. However, $\Delta$ is not DNlog for
  $r$ because the first formula of Definition~\ref{def-log-DN}
  does not hold.
  Indeed, as $\restrict{\tilde{X}}{\tau(p)}=X_1$,
  $\restrict{\tilde{Y}}{\tau(p)}=Y_1$ and $\mathcal{Y}=\{Y_1\}$,
  this formula is equivalent to
  $\mathcal{D}\models c \rightarrow \forall X_1[X_1\geq 0
  \rightarrow \exists Y_1 c]$.
  Let $v$ be a valuation such that
  $v(X_1)=v(Y_1)=v(X_2)=v(Y_2)=0$; then, $\mathcal{D}\models_v c$.
  Let $v_1$ be a valuation with $v_1(X_1)=1$ and $v_1$ matches $v$
  on the other variables; then,
  $\mathcal{D}\models_{v_1} X_1\geq 0$; however, 
  $\mathcal{D}\models_{v_1} \exists Y_1 c$ does not hold because $c$
  contains the constraint $X_2\geq X_1$ with $v_1(X_2)=0$ and
  $v_1(X_1)=1$ and it is not possible to change the value that $v_1$
  assigns to $Y_1$ so that $v_1(X_2)\geq v_1(X_1)$.
  Therefore, 
  $\mathcal{D}\models_v c \rightarrow \forall X_1[X_1\geq 0
  \rightarrow \exists Y_1 c]$ does not hold.
\end{example}
The point in Example~\ref{example-DN-implies-DNlog} is that
the problematic values (for DNlog-ness) cannot be captured by 
a query, hence they do not prevent $\Delta$ from being DN.
More precisely, we have
$p\big(v(X_1),v(X_2)\big)=p\big(v(Y_1),v(Y_2)\big)=p(0,0)$
and the atom $p(0,0)$ is captured by the query
$\query{p(0,0)}{\mathit{true}}$, \ie{}
$\Set{\query{p(0,0)}{\mathit{true}}}=\{p(0,0)\}$.
However, $p\big(v_1(X_1),v_1(X_2)\big)=p(1,0)$ and there exists no
query $Q$ with $\Set{Q}=\{p(1,0)\}$. 
If we had considered $r$ in the constraint domain $\mathcal{Q}_{\mathit{lin}}$
then $\Delta$ would not have been DN as there exists $Q_1$ such that
$\query{p(0,0)}{\mathit{true}}\lra_r Q_1$, the query $\query{p(1,0)}{\mathit{true}}$
is well-formed in $\mathcal{Q}_{\mathit{lin}}$ and is $\Delta$-more general
than $\query{p(0,0)}{\mathit{true}}$\footnote{because
$\restrict{\query{p(1,0)}{\mathit{true}}}{\overline{\tau}}
= \query{\restrict{p}{\overline{\tau}}(0)}{\mathit{true}} = 
\restrict{\query{p(0,0)}{\mathit{true}}}{\overline{\tau}}$ and
$\restrict{\query{p(1,0)}{\mathit{true}}}{\tau} =
\query{\restrict{p}{\tau}(1)}{\mathit{true}}$ with
$\Set{\query{\restrict{p}{\tau}(1)}{\mathit{true}}}=
\{\restrict{p}{\tau}(1)\}\subseteq\Set{\delta(p)}$},
but there exists no query $Q'_1$ such that
$\query{p(1,0)}{\mathit{true}}\lra_r Q'_1$.
Hence, an idea for matching DN with DNlog consists in considering
domains where every sequence of values can be captured by a query:
%
\begin{theorem}\label{prop-DN-implies-DNlog2}
  If, for all atoms $A$ whose arguments are elements of $D$, there exists
  a query $Q$ such that $\Set{Q}=\{A\}$,
  then every filter that is DN for a clause $r$ is also
  DNlog for $r$.
\end{theorem}
The intuition of the proof of Theorem~\ref{prop-DN-implies-DNlog2} consists
in mapping some sequences of values (induced by the considered valuations)
to queries that capture them and in using the DN property to prove
that DNlog-ness holds. More precisely, let
$r := p(\tilde{X}) \leftarrow c \diamond q(\tilde{Y})$
and $\Delta:=(\tau,\delta)$ be a filter that is DN for $r$.
First, we have to prove that
\[\mathcal{D} \models
c \rightarrow \forall_{\restrict{\tilde{X}}{\tau(p)}} \big[
\sat{\restrict{\tilde{X}}{\tau(p)}}{\delta(p)} \rightarrow
\exists_{\mathcal{Y}} c\big]\;.\]
Let $v$ be a valuation such that $\mathcal{D}\models_v c$ and
$v'$ be a valuation such that $v'(V)=v(V)$ for all variable 
$V\not\in \restrict{\tilde{X}}{\tau(p)}$ and 
$\mathcal{D}\models_{v'}\sat{\restrict{\tilde{X}}{\tau(p)}}{\delta(p)}$.
Then, there exists a query $Q$ such that $\Set{Q}=\{p([\tilde{X}]_v)\}$
and a query $Q'$ such that $\Set{Q'}=\{p([\tilde{X}]_{v'})\}$.
Intuitively, as $\mathcal{D}\models_v c$, there exists a derivation
step $Q\lra_r Q_1$; moreover, as $v'$ matches with $v$ on 
$\restrict{\tilde{X}}{\overline{\tau}(p)}$ and as the sequence of values
that $v'$ assignes to $\restrict{\tilde{X}}{\tau(p)}$ satisfies $\Delta$, then
$Q'$ is $\Delta$-more general than $Q$. Therefore, as $\Delta$ is DN for $r$,
there exists a query $Q'_1$ such that $Q'\lra_r Q'_1$ and $Q'_1$ is
$\Delta$-more general than $Q_1$; using these properties of $Q'$ and $Q'_1$,
one can deduce that $\mathcal{D}\models_{v'}\exists_{\mathcal{Y}}c$,
where $\mathcal{Y}=\restrict{\tilde{Y}}{\tau(q)}
\cup \mathit{local\_vars}(r)$. We also have to prove that
\[\mathcal{D}\models c\rightarrow
\sat{\restrict{\tilde{Y}}{\tau(q)}}{\delta(q)}\;.\]
This is a consequence of the fact that
for any derivation step $Q \lra_r Q_1$, the query $Q_1$ satisfies $\Delta$
(because $\Delta$ is DN for $r$).

\begin{example}
  \label{example-Qlin-DN-implies-DNlog}
  For any rational number $x$, there exists a term $t$ constructed
  from the constant and function symbols of $\mathcal{Q}_{\mathit{lin}}$
  such that $[t]_v=x$ for any valuation $v$. 
  Therefore, for each atom $p(\tilde{a})$ where $\tilde{a}$ is a sequence
  of rational numbers, there exists a query $Q$ in $\mathcal{Q}_{\mathit{lin}}$
  of the form $\query{p(\tilde{t})}{\mathit{true}}$, where
  the elements of $\tilde{t}$ are constructed
  from the constant and function symbols of $\mathcal{Q}_{\mathit{lin}}$,
  which is such that $\Set{Q}=\{p(\tilde{a})\}$. Hence,
  by Theorem~\ref{prop-DN-implies-DNlog2}, in $\mathcal{Q}_{\mathit{lin}}$
  DN is equivalent to DNlog. 
\end{example}

\subsection{Computing Looping Queries}
\label{section-computing-DN-filters}
%
For any filter $\Delta:=(\tau,\delta)$
and any clause $r:=p(\tilde{X})\leftarrow c \diamond q(\tilde{Y})$,
we let 
\begin{itemize}
\item $\mathrm{DNlog1}(\Delta,r) := \big(c \rightarrow
  \forall_{\restrict{\tilde{X}}{\tau(p)}} \big[
  \sat{\restrict{\tilde{X}}{\tau(p)}}{\delta(p)} \rightarrow
  \exists_{\mathcal{Y}} c\big]\big)$
\item $\mathrm{DNlog2}(\Delta,r) := \big(c\rightarrow
  \sat{\restrict{\tilde{Y}}{\tau(q)}}{\delta(q)}\big)$
\end{itemize}
denote the formulas in Definition~\ref{def-log-DN}.

A solution to compute a DNlog filter for a clause 
$r:=p(\tilde{X})\leftarrow c \diamond p(\tilde{Y})$
is to consider the \emph{projection} of $c$ on the elements
of $\tilde{X}$ that we wish to distinguish and to check that 
$\mathrm{DNlog1}$ and $\mathrm{DNlog2}$ hold for $r$
and the corresponding filter $\Delta_{\mathit{proj}}$. Formally,
for any set of variables $W$, the
projection of $c$ onto $W$ is denoted by ${\overline{\exists}_{W}c}$
and is the formula $\exists_{\Var(c)\setminus W}c$. If
$\mathrm{DNlog1}$ and $\mathrm{DNlog2}$ hold for $r$ and
$\Delta_{\mathit{proj}}$, then $\Delta_{\mathit{proj}}$ is
DNlog for $r$, hence it is DN for $r$ by
Theorem~\ref{prop-DNlog-implies-DN}; so we can try the test
of Theorem~\ref{propo-p-if-p-Delta} to get a query that loops
w.r.t. $\{r\}$. Hence the following algorithm:
%
\begin{center}
  \begin{tabular}{p{11.5cm}}
    \hline
    \multicolumn{1}{c}{An algorithm to compute a looping query}\\
    \hline
    Input: a clause $r:=p(\tilde{X})\leftarrow c \diamond p(\tilde{Y})$.
    \begin{enumerate}
    \item For each $m\subseteq [1,\arity{p}]$ do:
    \item \hspace{0.5cm}Set $\tau(p):=m$,
      $\delta(p):=\query{\restrict{p}{\tau}(\restrict{\tilde{X}}{\tau(p)})}
      {\overline{\exists}_{\restrict{\tilde{X}}{\tau(p)}}c}$
      and $\Delta_{\mathit{proj}}:=(\tau,\delta)$.
    \item \hspace{0.5cm}If $\mathrm{DNlog1}(\Delta_{\mathit{proj}},r)$ and
      $\mathrm{DNlog2}(\Delta_{\mathit{proj}},r)$ hold then 
    \item \hspace{1cm}If $\query{p(\tilde{Y})}{c}$ is
      $\Delta_{\mathit{proj}}$-more general
      than $\query{p(\tilde{X})}{c}$ then
    \item \hspace{1.5cm }return $\query{p(\tilde{X})}{c}$,
      which is a looping query w.r.t. $\{r\}$.
    \end{enumerate}\\[-2ex]
    \hline
  \end{tabular}
\end{center}
This algorithm \emph{always} finds a DNlog
filter. Indeed, for $m=\emptyset$, 
the corresponding filter $\Delta_{\mathit{proj}}=(\tau,\delta)$
is such that 
$\restrict{\tilde{X}}{\tau(p)}$ is the empty sequence, so
$\delta(p)=\query{\restrict{p}{\tau}}
{\overline{\exists}_{\emptyset}c}$ where 
$\overline{\exists}_{\emptyset}c$ is equivalent to
$\exists_{\Var(c)}c$ \ie{} to $\mathit{true}$ because in the
definition of a clause (see Section~\ref{sect-preliminaries}) we
suppose that $c$ is satisfiable; therefore, 
$\mathrm{DNlog1}(\Delta_{\mathit{proj}},r)$ and
$\mathrm{DNlog2}(\Delta_{\mathit{proj}},r)$ hold as they are equivalent
to
$c\rightarrow (\mathit{true}\rightarrow \exists_{\mathit{local\_vars(r)}}c)$ and
$c\rightarrow\mathit{true}$ respectively. 

Four tests are performed by the above
algorithm for each subset $m$ of $[1,\arity{p}]$: does
$\mathrm{DNlog1}(\Delta_{\mathit{proj}},r)$ hold and does
$\mathrm{DNlog2}(\Delta_{\mathit{proj}},r)$ hold
and, if these tests succeed, is 
$\restrict{\query{p(\tilde{Y})}{c}}{\overline{\tau}}$
more general than 
$\restrict{\query{p(\tilde{X})}{c}}{\overline{\tau}}$
and does $\query{p(\tilde{Y})}{c}$ satisfy $\Delta_{\mathit{proj}}$?
Actually, only three tests are necessary as we have:
%
\begin{lemma}
  \label{prop:DNlog2}
  Let $r:=p(\tilde{X})\leftarrow c \diamond p(\tilde{Y})$
  be a clause and $\Delta:=(\tau,\delta)$ be a filter.
  Then, we have $\mathcal{D}\models\mathrm{DNlog2}(\Delta,r)$ if and
  only if $\query{p(\tilde{Y})}{c}$ satisfies $\Delta$.
\end{lemma}
%
%

%
\begin{example}
  \label{ex-algo-works}
  Let us consider the constraint domain $\mathcal{Q}_{\mathit{lin}}$
  and the recursive clause
  \[r := p(X_1,X_2) \leftarrow X_1 \geq X_2 \land Y_1=X_1+1 \land Y_2=X_2
  \diamond p(Y_1,Y_2)\;.\]
  Let $c$ be the constraint in $r$. Consider $m:=\{1,2\}$.
  The projection of $c$ onto $\{X_1,X_2\}$
  is the constraint $X_1 \geq X_2$ hence the algorithm sets
  $\tau(p):=\{1,2\}$ and $\delta(p) := \query{p(X_1,X_2)}{X_1 \geq X_2}$
  and $\Delta_{\mathit{proj}}:=(\tau,\delta)$.
  The formulas $\mathrm{DNlog1}(\Delta_{\mathit{proj}},r)$ and
  $\mathrm{DNlog2}(\Delta_{\mathit{proj}},r)$ hold as they are respectively
  equivalent to
  \[c\rightarrow \forall X_1\forall X_2(X_1\geq X_2 \rightarrow
  \exists Y_1 \exists Y_2c)
  \qquad\textrm{and}\qquad
  c\rightarrow Y_1\geq Y_2\;.\]
  So, $\Delta_{\mathit{proj}}$
  is DNlog for $r$. Moreover, as $\query{p(Y_1,Y_2)}{c}$ is 
  $\Delta_{\mathit{proj}}$-more general than $\query{p(X_1,X_2)}{c}$, by 
  Theorem~\ref{propo-p-if-p-Delta} the query $\query{p(X_1,X_2)}{c}$
  loops w.r.t. $\{r\}$. Notice that by
  Definition~\ref{def-DN-filter}, every query that is $\Delta_{\mathit{proj}}$-more
  general than $\query{p(X_1,X_2)}{c}$ also loops w.r.t. $\{r\}$.
  Generally speaking, for any predicate symbol $q/n$, a set of positions
  $m\subseteq[1,n]$ can be seen as a finite representation of
  the set of queries of the form $\query{q(t_1,\dots,t_n)}{d}$ where
  for each $i \in m$, $d$ constrains $t_i$ to a ground term.
  For instance, $\query{p(0,0)}{\mathit{true}}$ loops w.r.t. $\{r\}$
  as it is $\Delta_{\mathit{proj}}$-more general than $\query{p(X_1,X_2)}{c}$;
  this query belongs to the class described by the set of positions
  $\{1,2\}$ for $p$; therefore we say that this class is non-terminating
  because \emph{there exists} a query in this class that loops.
  As $\query{p(0,X)}{\mathit{true}}$, $\query{p(X,0)}{\mathit{true}}$ and
  $\query{p(X,Y)}{\mathit{true}}$ are more general than
  $\query{p(0,0)}{\mathit{true}}$, by the
  Lifting Theorem~\ref{theorem-lifting} these queries also loop
  w.r.t. $\{r\}$; consequently, the classes described
  by the sets of positions $\{1\}$, $\{2\}$ and $\{\}$ for $p$
  are non-terminating too. So, for \emph{every} set of positions $m$
  for $p$, the class of queries described by
  $m$ is non-terminating.
\end{example}
%
%
\begin{example}
\label{ex-algo-fails}
  In $\mathcal{Q}_{\mathit{lin}}$ again, now consider the recursive clause
  (slightly different from that in Example~\ref{ex-algo-works})
  \[r := p(X_1,X_2) \leftarrow X_1 \leq X_2 \land Y_1=X_1+1 \land Y_2=X_2
  \diamond p(Y_1,Y_2)\]
  Let $c$ be the constraint in $r$ and $v$ be a valuation with
  $v(X_1)=v(X_2)=v(Y_2)=0$ and $v(Y_1)=1$; then we have
  $\mathcal{D}\models_v c$.
  \begin{itemize}
  \item Consider $m:=\{1,2\}$. The projection of $c$ onto $\{X_1,X_2\}$
    is $X_1 \leq X_2$ hence the algorithm sets
    $\tau(p):=\{1,2\}$, $\delta(p) := \query{p(X_1,X_2)}{X_1 \leq X_2}$
    and $\Delta_{\mathit{proj}}:=(\tau,\delta)$.
    The formula $\mathrm{DNlog2}(\Delta_{\mathit{proj}},r)$ is equivalent to
    $c\rightarrow Y_1\leq Y_2$. We have $\mathcal{D}\models_v c$ and
    $\mathcal{D}\not\models_v Y_1\leq Y_2$ so
    $\mathcal{D}\not\models_v c\rightarrow Y_1\leq Y_2$.
    Therefore, $\mathrm{DNlog2}(\Delta_{\mathit{proj}},r)$ does not hold,
    so $\Delta_{\mathit{proj}}$ is not DNlog for $r$.
  \item Consider $m:=\{1\}$. The projection of $c$ onto $\{X_1\}$ is equivalent
    to the constraint $\mathit{true}$. The algorithm sets $\tau(p):=\{1\}$,
    $\delta(p):=\query{\restrict{p}{\tau}(X_1)}{\mathit{true}}$ and
    $\Delta_{\mathit{proj}}:=(\tau,\delta)$. The formula
    $\mathrm{DNlog1}(\Delta_{\mathit{proj}},r)$ is equivalent to
    $c\rightarrow\forall X_1(\mathit{true}\rightarrow\exists Y_1 c)$
    \ie{}
    $c\rightarrow\forall X_1\exists Y_1 c$. 
    We have $\mathcal{D}\models_v c$; if we change the value
    assigned to $X_1$ to $1$, then $X_1\leq X_2$ (a subformula of $c$)
    does not hold anymore
    and one cannot find any value for $Y_1$ such that $X_1\leq X_2$
    holds again; therefore, we have
    $\mathcal{D}\not\models_v \forall X_1\exists Y_1 c$
    so
    $\mathcal{D}\not\models_v c\rightarrow\forall X_1\exists Y_1 c$.
    Hence, $\mathrm{DNlog1}(\Delta_{\mathit{proj}},r)$ does not hold, so
    $\Delta_{\mathit{proj}}$ is not DNlog for $r$. 
  \item Consider $m:=\{2\}$. The projection of $c$ onto $\{X_2\}$ is equivalent
    to the constraint $\mathit{true}$. The algorithm sets $\tau(p):=\{2\}$,
    $\delta(p):=\query{\restrict{p}{\tau}(X_2)}{\mathit{true}}$
    and $\Delta_{\mathit{proj}}:=(\tau,\delta)$. The formula
    $\mathrm{DNlog1}(\Delta_{\mathit{proj}},r)$ is equivalent to
    $c\rightarrow\forall X_2 (\mathit{true}\rightarrow\exists Y_2 c)$
    \ie{}
    $c\rightarrow\forall X_2 \exists Y_2 c$.
    We have $\mathcal{D}\models_v c$; if we change the value
    assigned to $X_2$ to $-1$, then $X_1\leq X_2$ (a subformula of $c$)
    does not hold anymore
    and one cannot find any value for $Y_2$ such that $X_1\leq X_2$
    holds again; therefore, we have
    $\mathcal{D}\not\models_v \forall X_2\exists Y_2 c$
    so
    $\mathcal{D}\not\models_v c\rightarrow\forall X_2\exists Y_2 c$.
    Hence, $\mathrm{DNlog1}(\Delta_{\mathit{proj}},r)$ does not hold, so
    $\Delta_{\mathit{proj}}$ is not DNlog for $r$. 
  \item Consider $m:=\emptyset$.
    The projection of $c$ onto $\emptyset$ is equivalent
    to the constraint $\mathit{true}$. The algorithm sets
    $\tau(p):=\emptyset$, $\delta(p):=\query{p_{\tau}}{\mathit{true}}$
    and $\Delta_{\mathit{proj}}:=(\tau,\delta)$.
    Both $\mathrm{DNlog1}(\Delta_{\mathit{proj}},r)$ and
    $\mathrm{DNlog2}(\Delta_{\mathit{proj}},r)$ hold as they are
    equivalent to
    $c\rightarrow (\mathit{true}\rightarrow c)$
    and $c\rightarrow \mathit{true}$ respectively.
    So, $\Delta_{\mathit{proj}}$ is DNlog for $r$.
    As $\query{p(Y_1,Y_2)}{c}$ is $\Delta_{\mathit{proj}}$-more general than
    $\query{p(X_1,X_2)}{c}$, by Theorem~\ref{propo-p-if-p-Delta}
    $\query{p(X_1,X_2)}{c}$ loops w.r.t. $\{r\}$.
    This query allows us to conclude that the class described by the set
    of positions $\{\}$ for $p$ is non-terminating.
  \end{itemize}
  Consequently, we get no information about the classes described
  by the sets of positions $\{1,2\}$, $\{1\}$ and $\{2\}$.
  Actually, the class described
  by $\{1,2\}$ is terminating, \ie{} \emph{every} query in this
  class does not loop; indeed, intuitively, when the arguments of
  $p$ in a query $Q$ are fixed to some values in $\Q$, we have a finite
  derivation of $\{r\}\cup\{Q\}$ because in $r$ the first argument of
  $p$ strictly increases until it becomes greater than the second argument.
  Hence, the class described by $\{1,2\}$ will not be inferred by our approach.  
  On the other hand, the query $\query{p(1,X)}{\mathit{true}}$ loops w.r.t.
  $\{r\}$, which implies that the class described by $\{1\}$ is
  non-terminating. Our approach fails to infer this result as  
  $X_1$ and $X_2$ interact in $c$ via $X_1\leq X_2$, so there is no DNlog filter
  for $r$ that distinguishes position 1 and not position 2 of $p$. Hence, as
  DN and DNlog match in this example, the DN approach fails%
  \footnote{Note that the situation of this example is different from that
    of Example~\ref{ex-algo-works}. Here, we cannot infer the non-termination
    of the class described by $\{1\}$ from the non-termination of the class described
    by $\{\}$. Indeed, every element in the class described by $\{1\}$ has the form
    $\query{p(t_1,t_2)}{d}$ where $d$ constrains $t_1$ to a ground term; on the
    other hand, every element in the class described by $\{\}$ has the form
    $\query{p(t'_1,t'_2)}{d'}$ where $t'_1$ and $t'_2$ are not constrained
    to some ground terms; hence $\query{p(t_1,t_2)}{d}$ is not more general
    than $\query{p(t'_1,t'_2)}{d'}$.} to infer the
  non-termination of $\{1\}$.
  So, a limitation of the DN approach when DN and DNlog match is the following:
  when two arguments interact,
  if there is no DNlog filter that distinguishes both their positions,
  then it is not possible to infer non-termination of a class of queries
  described by a set containing one of these positions and
  not the other.
  Notice that non-interaction
  of arguments is expressed by DNlog and not necessarily by DN; when DNlog and
  DN do not match (see Theorem~\ref{prop-DN-implies-DNlog2}), there are
  situations where DN arguments can interact with non-DN arguments. In
  Example~\ref{example-DN-implies-DNlog}, the arguments of $p$ at positions
  $1$ and $2$ interact via $X_2\geq X_1$; the filter 
  that we give in this example distinguishes position $1$ but not
  position $2$ of $p$ and it is DN for $r$.
\end{example}
\section{An Implementation}
\label{section-impl}

We have implemented the analysis in SWI-Prolog
\cite{Wielemaker:03b} for CLP($\mathcal{Q}_{lin}$). 
The prototype\footnote{available at
\texttt{http://personnel.univ-reunion.fr/fred/dev/DNlog4Q.zip}}
takes a recursive binary rule 
$p(\tilde{X}) \leftarrow c \diamond p(\tilde{Y})$  
as input and  tries to find a filter with
the projection of the constraint $c$ of the considered  rule onto
its head variables $\tilde{X}$. 
For each possible set of positions,  it computes the four logical
formulas corresponding to Definition~\ref{def-Delta-more-gen-state}
and Definition~\ref{def-log-DN}.  As the number of such sets is exponential
w.r.t. the arity of the predicate $p$,
our analysis is at least exponential.
These formulas are evaluated by a decision procedure for arbitrary logical
formulas over $\langle \Q ; \{0,1\} ; \{+\} ; \{=, <\} \rangle$. 
If they are
true (note that Lemma \ref{prop:DNlog2} shows 
that some tests are redundant), the analyzer prints the corresponding filter
and computes a concrete looping query. 

So the analyzer implements Theorem~\ref{propo-p-if-p-Delta} with the
help of Theorem \ref{prop-DNlog-implies-DN}.
We point out that the analysis can be automated for 
any constraint domain the theory of which is decidable, \eg \ logic programming
with finite trees and logic programming with rational trees \cite{Maher88}.

Table~1 summarizes the result of the analysis of a set of handcrafted binary rules.
The symbol $\checkmark$ indicates thoses examples that the analysis presented
in \cite{Payet04b} could not prove non-terminating.

\begin{table}
\caption{Running the analyzer on a set of examples.}
\begin{tabular}{lllll} \hline
\emph{binary clause} & $\tau$ & \multicolumn{1}{c}{$\delta$} & \emph{looping query} &  \\ \hline \hline
$p(A)  \leftarrow \mathit{true} \diamond p(B)$ & $\{ 1 \}$ & $\query{p(X)}{\mathit{true}}$ & $\query{p(0)}{\mathit{true}}$ &  \\ \hline
$p(A)  \leftarrow \mathit{A=B} \diamond p(B)$ & $\{ 1 \}$ & $\query{p(X)}{\mathit{true}}$ & $\query{p(0)}{\mathit{true}}$ &\\ \hline
$p(A)  \leftarrow A=0 \diamond p(B)$ & $\emptyset$ & $\query{p}{\mathit{true}}$ & $\query{p(A)}{A=0}$ &\\ \hline
$p(A)  \leftarrow A=0 \land B=0\diamond p(B)$ & $\emptyset$ & $\query{p}{\mathit{true}}$ & $\query{p(A)}{A=0}$ &\\ \hline
$p(A)  \leftarrow A=0\land B=1 \diamond p(B)$ &  &  & none found & \\ \hline
$p(A)  \leftarrow A\geq 0\land B=1 \diamond p(B)$ & $\{ 1 \}$ & $\query{p(X)}{X \geq 0}$ & $\query{p(0)}{\mathit{true}}$ &  \checkmark \\ \hline
$p(A)  \leftarrow A\geq 0\land B\geq1 \diamond p(B)$ & $\{ 1 \}$ & $\query{p(X)}{X \geq 0}$ & $\query{p(0)}{\mathit{true}}$ &  \checkmark\\ \hline
$p(A)  \leftarrow A\geq 0\land B\geq -1 \diamond p(B)$ & $\emptyset$ & $\query{p}{\mathit{true}}$ & $\query{p(A)}{A \geq 0}$ &\\ \hline
$p(A)  \leftarrow A\geq1 \land B\leq 0 \diamond p(B)$ &  &  & none found &\\ \hline
$p(A)  \leftarrow A=B+1 \land B\geq 0 \diamond p(B)$ & $\emptyset$ & $\query{p}{\mathit{true}}$ & $\query{p(A)}{A \geq 1}$&\\ \hline
$p(A,B)  \leftarrow A=C+1 \land C\geq 0$ & $\{ 2 \}$  &$\query{p(Y)}{\mathit{true}}$ & $\query{p(A,0)}{A \geq 1}$&\\
\hspace{16mm}$\diamond \ p(C,D)$ & &  & &\\ \hline
$p(A,B)  \leftarrow A=C+1 \land C\geq 0 $ & $\{ 2 \}$ & $\query{p(Y)}{\mathit{true}}$ & $\query{p(A,0)}{A \geq 1}$  &\\
\hspace{16mm}$\land B=D \diamond p(C,D)$ &  &  & &\\ \hline
$p(A,B)  \leftarrow A=C+1 \land C\geq 0$ & $\{ 2 \}$& $\query{p(Y)}{\mathit{true}}$&$\query{p(A,0)}{A \geq 1}$ &\\ 
\hspace{16mm}$\land B+1=D\diamond p(C,D)$ &  &  & &\\ \hline
$p(A,B)  \leftarrow A=C+1 \land C\geq 0 $& $\{ 2 \}$ & $\query{p(Y)}{Y \geq -1}$&$\query{p(A,-1)}{A \geq 1}$&  \checkmark\\
\hspace{16mm}$\land B+1=D \land D \geq 0$ & &  & &\\
\hspace{16mm}$\diamond\ p(C,D)$ & & & &\\ \hline
$p(A,B)  \leftarrow A=C+1 \land C\geq 0 $ & $\emptyset$ & $\query{p}{\mathit{true}}$ & $\langle p(A,B) | A \geq1$&\\
\hspace{16mm}$\land B=D+1 \land D \geq 0$ & & & $\land B \geq 1 \rangle$&\\
\hspace{16mm}$\diamond\; p(C,D)$ & & & &\\ \hline
$p(A,B) \leftarrow A\geq B \land C=A+1$ & \{1,2\} & $\query{p(X,Y)}{X \geq Y}$ & $\query{p(0,0)}{\mathit{true}}$&  \checkmark\\
\hspace{16mm} $\land D=BÊ\diamond p(C,D)$ & & & &\\ \hline                         
$p(A,B) \leftarrow A\leq B \land C=A+1$ & $\emptyset$ & $\query{p}{\mathit{true}}$ & $\query{p(A,B)}{A \leq B}$ &\\
\hspace{16mm} $\land D=B \diamond p(C,D)$ & & & &\\ \hline
$pow2(A,B,C)  \leftarrow $ & $\{ 2, 3 \}$ & $\langle pow2(Y,Z) |$ & $\langle pow2(A,1,2) |$&  \checkmark\\
\hspace{16mm}$A=D+1 \land D \geq 0$ & & $Y \geq1 \land Z \geq 2\rangle $ & $A \geq 1\rangle $&\\
\hspace{16mm}$\land E=2*B \land B \geq 1$ & & & &\\
\hspace{16mm}$\land F=C\land C \geq 2$ & &  & & \\
\hspace{16mm}$\diamond\ pow2(D,E,F)$ & & & &\\ \hline
\end{tabular}
\vspace{-2\baselineskip}
\end{table}

\section{Conclusion}
\label{section-conclusion}
In~\cite{Payet06a} we have presented a technique to complement termination
analysis with non-termination inside the logic programming paradigm. Our
aim was to detect optimal termination conditions expressed in a language
describing classes of queries. The approach 
was syntactic and 
linked to some basic logic programming machinery such as
the unification algorithm.
In~\cite{Payet04b} we have presented a first step at generalizing
the work of~\cite{Payet06a} to the CLP setting. The logical
criterion we gave only considers those filters,
the function $\delta$ of which does not filter anything \ie{}
$\delta$ maps any predicate symbol $p$ to
$\query{\restrict{p}{\tau}(\tilde{X})}{\mathit{true}}$.
 
This paper describes a generalization of~\cite{Payet06a}
to the CLP setting. It presents a criterion, both in an operational
and a logical form, to infer non-terminating
atomic queries with respect to a binary CLP clause.
This criterion is generic in the constraint domain;
its logical form strictly generalizes that of~\cite{Payet04b}
and it has been fully implemented for CLP($\mathcal{Q}_{lin}$).

\section*{Acknowledgments}
The authors thank the anonymous reviewers for helpful comments on 
the previous versions of this paper.

\bibliographystyle{acmtrans}

\appendix
\newtheorem{lemma-app}{Lemma}
\newtheorem{proposition-app}[lemma-app]{Proposition}

\section{-- Proof of the results in
  Section~\ref {section-loop-inference-with-constraints}}
\label{appendixSect4}

\subsection{-- Lemma~\ref{lemma-set-variant}}
If $\Set{Q}=\emptyset$ then $\Set{Q}\subseteq\Set{Q'}$.
Otherwise, let $\query{p(\tilde{t})}{d}:=Q$ and let
$\query{p(\tilde{t'})}{d'}:=Q'$.
As $Q'$ is a variant of $Q$, there exists a renaming $\gamma$
such that $\tilde{t'}=\gamma(\tilde{t})$
and $d'=\gamma(d)$.  
Let $p(\tilde{a})\in\Set{Q}$. Then, there exists
a valuation $v$ such that $\tilde{a}=[\tilde{t}]_v$ and
$\mathcal{D} \models_v d$. Let $v_1$ be the
valuation defined as: for all variable $V$,
$v_1(V)=v(\gamma^{-1}(V))$. Then, we have
$[\tilde{t'}]_{v_1} = [\gamma(\tilde{t})]_{v_1} =
[\gamma^{-1}(\gamma(\tilde{t}))]_v = [\tilde{t}]_v = \tilde{a}$.
Moreover, $[d']_{v_1}=[\gamma(d)]_{v_1}=[\gamma^{-1}(\gamma(d))]_v
=[d]_v=1$. Consequently, $\mathcal{D} \models_{v_1} d'$.
Therefore, $p(\tilde{a})\in\Set{Q'}$.

So, we always have $\Set{Q}\subseteq\Set{Q'}$. The proof of
$\Set{Q'}\subseteq\Set{Q}$ follows by symmetry.

\subsection{-- Lemma~\ref{lemma-operational-sem}}
Let $\query{p(\tilde{u})}{d}:=Q$.
\begin{itemize}
\item[$\Rightarrow$)] Suppose that there exists a derivation step
  of the form $Q\lra_r Q_1$. Then, $H$ has the form $p(\tilde{s})$.
  Let $r':=p(\tilde{s}') \leftarrow c'\diamond B'$
  be the input clause of this step. We have
  $\mathcal{D} \models \exists (\tilde{u}=\tilde{s}'\land
  c'\land d)$. So, there exists a valuation $v$ such that
  $\mathcal{D} \models_v (\tilde{u}=\tilde{s}'\land
  c'\land d)$. Notice that:
  \begin{center}
    $p([\tilde{u}]_v)\in\Set{Q}$ and
    $p([\tilde{s}']_v)\in\Set{\query{p(\tilde{s}')}{c'}}$ and
    $[\tilde{u}]_v = [\tilde{s}']_v$.
  \end{center}
  Hence, $\Set{Q}\cap\Set{\query{p(\tilde{s}')}{c'}} \neq\emptyset$.
  As $\query{p(\tilde{s}')}{c'}$ is a variant of
  $\query{p(\tilde{s})}{c}$, by Lemma~\ref{lemma-set-variant}
  we have 
  $\Set{Q}\cap\Set{\query{p(\tilde{s})}{c}} \neq\emptyset$
  \ie{} $\Set{Q}\cap\Set{\query{H}{c}} \neq\emptyset$.
\item[$\Leftarrow$)] Suppose that
  $\Set{Q}\cap\Set{\query{H}{c}} \neq\emptyset$.
  Then, $H$ has the form $p(\tilde{s})$ and we have
  $\Set{Q}\cap\Set{\query{p(\tilde{s})}{c}} \neq\emptyset$.
  Let $r':=p(\tilde{s}') \leftarrow c'\diamond B'$ be a variant of $r$
  variable disjoint with $Q$. By Lemma~\ref{lemma-set-variant},
  $\Set{\query{p(\tilde{s}')}{c'}}=\Set{\query{p(\tilde{s})}{c}}$,
  so we have
  $\Set{Q}\cap\Set{\query{p(\tilde{s}')}{c'}} \neq\emptyset$.
  Let $p(\tilde{a})\in\Set{Q}\cap\Set{\query{p(\tilde{s}')}{c'}}$.
  Then, there exists:
  \begin{itemize}
  \item a valuation $v_1$ such that $\tilde{a}=[\tilde{u}]_{v_1}$
    and $\mathcal{D} \models_{v_1} d$,
  \item a valuation $v_2$ such that $\tilde{a}=[\tilde{s'}]_{v_2}$
    and $\mathcal{D} \models_{v_2} c'$.
  \end{itemize}
  As $r'$ and $Q$ are variable disjoint, there exists a valuation $v$
  such that:
  \begin{itemize}
  \item for all variable $V\in\Var(Q)$, $v(V)=v_1(V)$ and
  \item for all variable $V\in\Var(r')$, $v(V)=v_2(V)$.
  \end{itemize}
  Then, we have
  $[\tilde{u}]_v = [\tilde{u}]_{v_1} = \tilde{a}$,
  $[\tilde{s}']_v = [\tilde{s}']_{v_2} = \tilde{a}$,
  $[d]_v=[d]_{v_1}=1$ and
  $[c']_v=[c']_{v_2}=1$.
  Consequently, $\mathcal{D} \models_v
  (\tilde{u}=\tilde{s}' \land c'\land d)$. Hence,
  $\mathit{solv}(\tilde{u}=\tilde{s}' \land c'\land d)=\mathtt{true}$,
  so we have
  $Q\lra_r\query{B'}{\tilde{u}=\tilde{s}' \land c'\land d}$.
\end{itemize}

\subsection{-- Theorem~\ref{theorem-lifting}}
We have already proved that there exists a query $Q'_1$ such that
$Q'\lra_r Q'_1$ (see beginning of Section~\ref{section-loop-constraints}).
Let $\query{p(\tilde{u})}{d}:=Q$ and $\query{p(\tilde{u}')}{d'}:=Q'$.
Let $r_1:=p(\tilde{s}_1)\leftarrow c_1\diamond q(\tilde{t}_1)$
be the input clause in $Q\lra_r Q_1$ and
$r'_1:=p(\tilde{s}'_1)\leftarrow c'_1\diamond q(\tilde{t}'_1)$
be the input clause in $Q'\lra_r Q'_1$. Then,
\[Q_1 = \query{q(\tilde{t}_1)}{\tilde{u}=\tilde{s}_1\land c_1\land d}
\quad \text{and} \quad
Q'_1=\query{q(\tilde{t}'_1)}{\tilde{u}'=\tilde{s}'_1\land c'_1\land d'}\;.\]
Let us prove that $Q'_1$ is more general than $Q_1$ \ie{} that
$\Set{Q_1}\subseteq\Set{Q'_1}$.
If $\Set{Q_1}$ is empty, then the result trivially holds. Suppose that
$\Set{Q_1}$ is not empty.
Let $q(\tilde{a})\in\Set{Q_1}$. Then, there exists a valuation $v$ such that
\begin{equation}\label{proof-theorem-lifting-eq1}
  \tilde{a}=[\tilde{t}_1]_v \quad \text{and}\quad 
  \mathcal{D} \models_v
  (\tilde{u}=\tilde{s}_1\land c_1\land d)\;.
\end{equation}
Hence, $\mathcal{D} \models_v d$, so
$p([\tilde{u}]_v)\in\Set{Q}$. As $Q'$ is more general than $Q$, then
$p([\tilde{u}]_v)\in\Set{Q'}$. Consequently, there exists a valuation $v'_1$
such that
\begin{equation}\label{proof-theorem-lifting-eq2}
  [\tilde{u}]_v = [\tilde{u}']_{v'_1} \quad \text{and} \quad
  \mathcal{D} \models_{v'_1} d'\;.
\end{equation}
Notice that $r_1$ and $r'_1$ are variants, so $r_1=\gamma(r'_1)$
for a renaming $\gamma$. As $Q'$ and
$r'_1$ are variable disjoint (because $r'_1$ is the input clause in
$Q'\lra_r Q'_1$), there exists a valuation $v'$ such that:
\begin{itemize}
\item for all variable $V\in\Var(r'_1)$, $v'(V)=v(\gamma(V))$ and 
\item for all variable $V\in\Var(Q')$, $v'(V)=v'_1(V)$. 
\end{itemize}
Then, we have 
$[\tilde{s}'_1]_{v'} = [\gamma(\tilde{s}'_1)]_v = [\tilde{s}_1]_v$
with $[\tilde{s}_1]_v = [\tilde{u}]_v$
by~(\ref{proof-theorem-lifting-eq1}) and
$[\tilde{u}']_{v'} = [\tilde{u}']_{v'_1}$ with
$[\tilde{u}']_{v'_1} = [\tilde{u}]_v$
by~(\ref{proof-theorem-lifting-eq2}). So,
$[\tilde{s}'_1]_{v'}=[\tilde{u}']_{v'}$.
Moreover,
$[c'_1]_{v'}=[\gamma(c'_1)]_v=[c_1]_v$ with
$[c_1]_v=1$ by~(\ref{proof-theorem-lifting-eq1}) and
$[d']_{v'}=[d']_{v'_1}$ with $[d']_{v'_1}=1$
by~(\ref{proof-theorem-lifting-eq2}).
So, we have
$\mathcal{D} \models_{v'}
(\tilde{u}'=\tilde{s}'_1\land c'_1\land d')$.
As $[\tilde{t}'_1]_{v'} = [\gamma(\tilde{t}'_1)]_v =
[\tilde{t}_1]_v$ with $[\tilde{t}_1]_v = \tilde{a}$
by~(\ref{proof-theorem-lifting-eq1}), we conclude that
$q(\tilde{a})\in\Set{Q'_1}$.

\subsection{-- Corollary~\ref{coro-p-if-p} and Corollary~\ref{coro-p-if-q}}
First, we need a lemma.
\begin{lemma-app}\label{lemma-useful-loop}
  Let $r := H \leftarrow c \diamond B$ be a clause. Then,
  there exists a derivation step $\query{H}{c}\lra_r Q$ where
  $\Set{Q}=\Set{\query{B}{c}}$.
\end{lemma-app}
\begin{proof}
  As $\mathcal{D} \models \exists c$
  (by definition of a clause), we have $\Set{\query{H}{c}}\neq \emptyset$.
  Hence, $\Set{\query{H}{c}}\cap\Set{\query{H}{c}}\neq \emptyset$.
  Consequently, by Lemma~\ref{lemma-operational-sem},
  there exists a derivation step of the form $\query{H}{c}\lra_r Q$.
  Let us prove that $\Set{Q}=\Set{\query{B}{c}}$.
  Let $p(\tilde{s}):=H$ and $q(\tilde{t}):=B$.
  Let $r' := p(\tilde{s}') \leftarrow c' \diamond q(\tilde{t}')$
  be the input clause in $\query{H}{c}\lra_r Q$. Then,
  $Q=\query{q(\tilde{t}')}{\tilde{s}'=\tilde{s} \land c' \land c}$.
  Let $\gamma$ be a renaming such that $r=\gamma(r')$.
  \begin{itemize}
  \item Let us prove that $\Set{\query{B}{c}}\subseteq\Set{Q}$.
    If $\Set{\query{B}{c}}$ is empty, then the result holds. Suppose that
    $\Set{\query{B}{c}}$ is not empty.
    Let $q(\tilde{a})\in\Set{\query{B}{c}}$.
    Then, there exists a valuation $v$ such that
    $\tilde{a}=[\tilde{t}]_v$ and $\mathcal{D} \models_v c$.
    Let $v_1$ be the valuation defined as:
    \begin{itemize}
    \item for all variable $V\in\Var(r')$, $v_1(V)=v(\gamma(V))$ and
    \item for all variable $V\not\in\Var(r')$, $v_1(V)=v(V)$.
    \end{itemize}
    Then, we have
    $[\tilde{t}']_{v_1} = [\gamma(\tilde{t}')]_v =
    [\tilde{t}]_v = \tilde{a}$ and
    $[\tilde{s}']_{v_1} = [\gamma(\tilde{s}')]_v = [\tilde{s}]_v$ with
    $[\tilde{s}]_v = [\tilde{s}]_{v_1}$ because, as $r'$ is the input clause
    in $\query{H}{c}\lra_r Q$, $r'$ is variable disjoint with
    $\query{H}{c}=\query{p(\tilde{s})}{c}$. Moreover,
    $[c']_{v_1}=[\gamma(c')]_v=[c]_v=1$ and
    $[c]_{v_1}=[c]_v$ (because $r'$ is variable disjoint with
    $\query{p(\tilde{s})}{c}$) \ie{} $[c]_{v_1}=1$.
    Consequently, $[\tilde{t}']_{v_1} = \tilde{a}$ and
    $\mathcal{D} \models_{v_1}
    (\tilde{s}=\tilde{s}' \land c' \land c)$. Hence,
    $q(\tilde{a})\in\Set{Q}$.
  \item Let us prove that $\Set{Q}\subseteq\Set{\query{B}{c}}$.
    If $\Set{Q}$ is empty, then the result holds. Suppose that
    $\Set{Q}$ is not empty.
    Let $q(\tilde{a})\in\Set{Q}$.
    Then, there exists a valuation $v$ such that
    $\tilde{a}=[\tilde{t}']_v$ and
    $\mathcal{D} \models_v (\tilde{s}'=\tilde{s} \land c' \land c)$.
    Let $v_1$ be a valuation such that:
    for all variable $V\in\Var(r)$, $v_1(V)=v(\gamma^{-1}(V))$.
    Then, we have
    $[\tilde{t}]_{v_1} = [\gamma^{-1}(\tilde{t})]_v =
    [\tilde{t}']_v = \tilde{a}$ and
    $[c]_{v_1}=[\gamma^{-1}(c)]_v=[c']_v=1$.
    Consequently, $[\tilde{t}]_{v_1} = \tilde{a}$ and
    $\mathcal{D} \models_{v_1} c$.
    Hence, $q(\tilde{a})\in\Set{\query{B}{c}}$. 
  \end{itemize}
\end{proof}

\begin{proof}[Corollary~\ref{coro-p-if-p}]
  By Lemma~\ref{lemma-useful-loop}, there exists a derivation
  step of the form $\query{H}{c} \lra_r Q$
  with $\Set{Q}=\Set{\query{B}{c}}$. Then,
  $\Set{\query{H}{c}}\subseteq\Set{Q}$ (because
  $\Set{\query{H}{c}}\subseteq\Set{\query{B}{c}}$)
  so, by repeatedly using the Lifting
  Theorem~\ref{theorem-lifting}, one can
  build an infinite derivation of
  $\{r\}\cup\{\query{H}{c}\}$. Consequently, $\query{H}{c}$
  loops w.r.t. $\{r\}$. 
\end{proof}

\begin{proof}[Corollary~\ref{coro-p-if-q}]
  By Lemma~\ref{lemma-useful-loop}, we have
  $\query{H}{c} \lra_r Q$
  where $Q$ is more general than $\query{B}{c}$.
  As there exists an infinite derivation $\xi$ of
  $P\cup\{\query{B}{c}\}$,
  by successively applying the Lifting
  Theorem~\ref{theorem-lifting}
  to each step of $\xi$ one can construct an infinite
  derivation of $P\cup\{Q\}$. Consequently, $\query{H}{c}$
  loops w.r.t. $P$. 
\end{proof}

\newpage
\section{-- Proof of the results in
  Section~\ref {section-loop-filters}}
\label{appendixSect5}

\subsection{-- Lemma~\ref{lemma-restriction}
  and Lemma~\ref{lemma-useful2-theo-DNlog-iff-DN}}
%
\begin{proof}[Lemma~\ref{lemma-restriction}]
  If $\Set{Q}=\emptyset$ then $\Set{\restrict{Q}{\tau}}=\emptyset$,
  so the result holds.
  Otherwise, as $\Set{Q}\subseteq\Set{Q'}$, then $\rel{Q}=\rel{Q'}$ \ie{}
  $Q$ has the form $\query{p(\tilde{t})}{d}$ and $Q'$
  has the form $\query{p(\tilde{t'})}{d'}$.
  Notice that
  \[\restrict{Q}{\tau}=\query{\restrict{p}{\tau}
    (\restrict{\tilde{t}}{\tau(p)})}{d}
  \quad\text{and}\quad
  \restrict{Q'}{\tau}=
  \query{\restrict{p}{\tau}(\restrict{\tilde{t}'}{\tau(p)})}{d'}\;.\]
  If $\Set{\restrict{Q}{\tau}}$ is empty, then the result holds. 
  Suppose that $\Set{\restrict{Q}{\tau}}$ is not empty.
  Let $\restrict{p}{\tau}(\tilde{a})\in\Set{\restrict{Q}{\tau}}$.
  Then, there exists a valuation
  $v$ such that $\tilde{a}=[\restrict{\tilde{t}}{\tau(p)}]_v$ and
  $\mathcal{D} \models_v d$. Let $\tilde{b}$ be the sequence
  of $\arity{p}$ elements of $D$ defined as:
  \begin{itemize}
  \item $\restrict{\tilde{b}}{\tau(p)}=\tilde{a}$, \ie{}
    $\restrict{\tilde{b}}{\tau(p)} =
    [\restrict{\tilde{t}}{\tau(p)}]_v$, and
  \item $\restrict{\tilde{b}}{\overline{\tau}(p)}=
    [\restrict{\tilde{t}}{\overline{\tau}(p)}]_v$.
  \end{itemize}
  Then, we have $\tilde{b}=[\tilde{t}]_v$ with
  $\mathcal{D} \models_v d$.
  Therefore, $p(\tilde{b})\in\Set{Q}$. As $\Set{Q}\subseteq\Set{Q'}$,
  then $p(\tilde{b})\in\Set{Q'}$. Consequently, there exists a valuation
  $v'$ such that $\tilde{b}=[\tilde{t'}]_{v'}$ and
  $\mathcal{D} \models_{v'} d'$.
  Hence, we have
  $\tilde{a}=\restrict{\tilde{b}}{\tau(p)}=
  [\restrict{\tilde{t}'}{\tau(p)}]_{v'}$
  and $\mathcal{D} \models_{v'} d'$.
  So, $\restrict{p}{\tau}(\tilde{a})\in\Set{\restrict{Q'}{\tau}}$. 
\end{proof}

\begin{proof}[Lemma~\ref{lemma-useful2-theo-DNlog-iff-DN}]
  If $\Set{Q}\cap\Set{Q'}\neq\emptyset$
  then there exists $p(\tilde{a})\in\Set{Q}\cap\Set{Q'}$ \ie{}
  $p(\tilde{a})\in\Set{Q}$ and
  $p(\tilde{a})\in\Set{Q'}$. This implies that
  $\restrict{p}{\tau}(\restrict{\tilde{a}}{\tau(p)})\in
  \Set{\restrict{Q}{\tau}}$ and
  $\restrict{p}{\tau}(\restrict{\tilde{a}}{\tau(p)})\in
  \Set{\restrict{Q'}{\tau}}$. So,
  $\restrict{p}{\tau}(\restrict{\tilde{a}}{\tau(p)})\in
  \Set{\restrict{Q}{\tau}}\cap\Set{\restrict{Q'}{\tau}}$.
  Therefore, $\Set{\restrict{Q}{\tau}}\cap
  \Set{\restrict{Q'}{\tau}}\neq\emptyset$.
\end{proof}

\subsection{-- Lemma~\ref{lemma-properties-delta}}
Let $\Delta:=(\tau,\delta)$ be a filter.
Let $Q$, $Q'$ and $Q{''}$
be some queries such that $Q{''}$ is $\Delta$-more
general than $Q'$ and $Q'$ is $\Delta$-more general
than $Q$. 
As $Q{''}$ is $\Delta$-more general than $Q'$, then
$Q^{''}_{\overline{\tau}}$ is more general than
$Q'_{\overline{\tau}}$ and
$Q{''}$ satisfies $\Delta$.
As $Q{'}$ is $\Delta$-more general than $Q$, then
$Q'_{\overline{\tau}}$ is more general than
$Q_{\overline{\tau}}$.
Consequently,
$Q^{''}_{\overline{\tau}}$ is more general than
$Q_{\overline{\tau}}$ (because
the ``more general than'' relation is transitive) and
$Q{''}$ satisfies $\Delta$.
Therefore, $Q{''}$ is $\Delta$-more general than $Q$.

\subsection{-- Theorem~\ref{propo-p-if-p-Delta}}
By Lemma~\ref{lemma-useful-loop}, we have
$\query{H}{c}\lra_r Q$ where $\Set{Q}=\Set{\query{B}{c}}$.
So by Lemma~\ref{lemma-restriction},
\begin{center}
  $\restrict{Q}{\overline{\tau}}$
  is more general than $\restrict{\query{B}{c}}{\overline{\tau}}$
\end{center}
and
$\Set{\restrict{Q}{\tau}}\subseteq\Set{\restrict{\query{B}{c}}{\tau}}$.
As $\query{B}{c}$ satisfies $\Delta$ (because $\query{B}{c}$ is
$\Delta$-more general than $\query{H}{c}$), we have
$\Set{\restrict{\query{B}{c}}{\tau}}\subseteq\Set{\delta(q)}$ where
we let $q:=\rel{B}$. Hence, $\Set{\restrict{Q}{\tau}}\subseteq\Set{\delta(q)}$
\ie{}
\begin{center}
  $Q$ satisfies $\Delta$.
\end{center}
Therefore, $Q$ is $\Delta$-more general than $\query{B}{c}$.
So, as $\query{B}{c}$ is $\Delta$-more general than $\query{H}{c}$
and the ``$\Delta$-more general than'' relation is transitive (by
Lemma~\ref{lemma-properties-delta}), we have that
$Q$ is $\Delta$-more general than $\query{H}{c}$.
As $\Delta$ is DN for $r$, by repeatedly using Definition~\ref{def-DN-filter},
one can build an infinite derivation of $\{r\}\cup\{\query{H}{c}\}$. 
Consequently, $\query{H}{c}$ loops w.r.t. $\{r\}$.

\subsection{-- Lemma~\ref{lemma-sat-set}}
Let $\query{p(\tilde{s})}{d}:=Q$.
Let $Q':=\query{p(\tilde{s}')}{d'}$ be a variant of $Q$ variable
disjoint with $\tilde{u}$.
\begin{itemize}
\item[$\Rightarrow$)]
  Suppose that $p([\tilde{u}]_v)\in\Set{Q}$. Then, as by
  Lemma~\ref{lemma-set-variant} $\Set{Q}=\Set{Q'}$,
  we have $p([\tilde{u}]_v)\in\Set{Q'}$. Hence, there
  exists a valuation $w$ such that
  $[\tilde{u}]_v = [\tilde{s}']_w$ and
  $\mathcal{D} \models_w d'$.    
  Let $v_1$ be a valuation such that:
  \begin{itemize}
  \item for all variable $V\in\Var(Q')$, 
    $v_1(V)=w(V)$ and
  \item for all variable $V\not\in\Var(Q')$, 
    $v_1(V)=v(V)$.
  \end{itemize}
  Then, as $Q'$ and $\tilde{u}$ are variable disjoint,
  $[\tilde{u}]_{v_1} = [\tilde{u}]_v$. Moreover,
  $[\tilde{s}']_{v_1} = [\tilde{s}']_w = [\tilde{u}]_v$
  and $[d']_{v_1}=[d']_w=1$.
  Hence, $\mathcal{D} \models_{v_1}
  (\tilde{u} = \tilde{s}' \land d')$. Therefore,
  $\mathcal{D} \models_v
  \exists_{\Var(Q')}
  (\tilde{u} = \tilde{s}' \land d')$ \ie{}
  $\mathcal{D} \models_v
  \sat{\tilde{u}}{Q}$.
\item[$\Leftarrow$)]
  Suppose that $\mathcal{D} \models_v
  \sat{\tilde{u}}{Q}$ \ie{}
  $\mathcal{D} \models_v
  \exists_{\Var(Q')}
  (\tilde{u} = \tilde{s}' \land d')$.
  Then, there exists a valuation $v_1$ such that
  \begin{itemize}
  \item $\mathcal{D} \models_{v_1}
    (\tilde{u} = \tilde{s}' \land d')$ and
  \item for all variable $V\not\in\Var(Q')$, 
    $v_1(V)=v(V)$.
  \end{itemize}
  As $Q'$ and $\tilde{u}$ are variable disjoint,
  we have $[\tilde{u}]_v = [\tilde{u}]_{v_1}$.
  Moreover, $[\tilde{u}]_{v_1} = [\tilde{s}']_{v_1}$ and
  $\mathcal{D} \models_{v_1} d'$.
  Consequently, $p([\tilde{u}]_v)\in\Set{Q'}$.
  As, by Lemma~\ref{lemma-set-variant}, $\Set{Q}=\Set{Q'}$,
  we have $p([\tilde{u}]_v)\in\Set{Q}$.
\end{itemize}

\subsection{-- Theorem~\ref{prop-DNlog-implies-DN}}
%
%
First, we need a technical lemma:
%
\begin{lemma-app}\label{lemma-useful-theo-DNlog-iff-DN}
  Let $Q:=\query{p(\tilde{u})}{d}$ and $Q':=\query{p(\tilde{u}')}{d'}$
  be two variable disjoint queries. If $\Set{Q}\cap\Set{Q'}\neq\emptyset$
  then there exists a valuation $v$ such that
  $\mathcal{D} \models_v (\tilde{u}=\tilde{u}'\land d\land d')$.
\end{lemma-app}
\begin{proof}
  Suppose that $\Set{Q}\cap\Set{Q'}\neq\emptyset$. Then, there
  exists $p(\tilde{a})$ such that $p(\tilde{a})\in\Set{Q}$ and
  $p(\tilde{a})\in\Set{Q'}$. Hence, there exists:
  \begin{itemize}
  \item a valuation $v_1$ such that $\tilde{a}=[\tilde{u}]_{v_1}$
    and $\mathcal{D} \models_{v_1} d$ and
  \item a valuation $v_2$ such that $\tilde{a}=[\tilde{u}']_{v_2}$
    and $\mathcal{D} \models_{v_2} d'$.
  \end{itemize}
  As $Q$ and $Q'$ are variable disjoint, there exists a valuation
  $v$ such that:
  \begin{itemize}
  \item for all variable $V\in\Var(Q)$, $v(V)=v_1(V)$ and
  \item for all variable $V\in\Var(Q')$, $v(V)=v_2(V)$.
  \end{itemize}
  Then, $[d]_v=[d]_{v_1}=1$, $[d']_v=[d']_{v_2}=1$ and
  $[\tilde{u}]_v = [\tilde{u}]_{v_1} = \tilde{a} =
  [\tilde{u}']_{v_2} = [\tilde{u}']_v$.
  Consequently, $\mathcal{D} \models_v
  (\tilde{u}=\tilde{u}'\land d\land d')$.
\end{proof}

Given a clause $r$ and a filter $\Delta$ that is DNlog for $r$, we
have to prove that $\Delta$ is DN for $r$.
By Definition~\ref{def-DN-filter}, given a derivation step
$Q \lra_r T$, we have to establish the following facts:
\begin{itemize}
\item[Fact 1.]
  The query $T$ satisfies $\Delta$.
\item[Fact 2.]
  For each query $Q'$ that is $\Delta$-more general than $Q$,
  there exists a derivation step $Q'\lra_r T'$ 
  where $T'$ is $\Delta$-more general than $T$.
\end{itemize}
Fact~1 is established by Proposition~\ref{prop-DNlog-implies-DN-1} below.
We prove Fact~2 in two steps; given a query $Q'$ that is
$\Delta$-more general than $Q$, we prove that: 
\begin{itemize}
\item[Fact 2a.]
  there exists a derivation step $Q'\lra_r T'$ where $T'$ satisfies
  $\Delta$ (see Proposition~\ref{prop-DNlog-implies-DN-2} below).
\item[Fact 2b.]
  the query $T'$ in $Q'\lra_r T'$ is such that
  $\restrict{T'}{\overline{\tau}}$
  is more general than $\restrict{T}{\overline{\tau}}$
  (see Proposition~\ref{prop-DNlog-implies-DN-3} below).
\end{itemize}
Then by Definition~\ref{def-Delta-more-gen-state}, the query $T'$ is
$\Delta$-more general than $T$.

\begin{proposition-app}\label{prop-DNlog-implies-DN-1}
  Let $\Delta$ be a filter that is DNlog for a clause $r$ and
  $Q\lra_r T$ be a derivation step. Then, $T$ satisfies $\Delta$.
\end{proposition-app}
\begin{proof}
  Let $(\tau,\delta) := \Delta$ and
  $\query{p(\tilde{u})}{d}:=Q$. Let
  $r_1 := p(\tilde{X})\la c\diamond q(\tilde{Y})$ be the
  input clause in $Q\lra_r T$. Then,
  $T = \query{q(\tilde{Y})}{\tilde{X}=\tilde{u} \land c \land d}$.
  Let us prove that $T$ satisfies $\Delta$ \ie{} that
  $\Set{\restrict{T}{\tau}}\subseteq\Set{\delta(q)}$.
  Let $\restrict{q}{\tau}(\tilde{a})\in\Set{\restrict{T}{\tau}}$.
  Then, there exists a valuation $v$ such that
  \begin{equation}\label{prop-DNlog-implies-DN-1-eq1}
    \tilde{a}=[\restrict{\tilde{Y}}{\tau(q)}]_v
    \quad\text{and}\quad
    \mathcal{D}\models_v\tilde{X}=\tilde{u}\land c \land d~.
  \end{equation}
  As $\Delta$ is DNlog for $r$,
  it is also DNlog for $r_1$. Consequently, we have
  $\mathcal{D}\models_v c\rightarrow
  \sat{\restrict{\tilde{Y}}{\tau(q)}}{\delta(q)}$.
  As $\mathcal{D}\models_v c$ (by~(\ref{prop-DNlog-implies-DN-1-eq1})),
  then we have $\mathcal{D}\models_v
  \sat{\restrict{\tilde{Y}}{\tau(q)}}{\delta(q)}$.
  Therefore, by Lemma~\ref{lemma-sat-set},
  $\restrict{q}{\tau}([\restrict{\tilde{Y}}{\tau(q)}]_v)
  \in \Set{\delta(q)}$ \ie{}
  $\restrict{q}{\tau}(\tilde{a})\in\Set{\delta(q)}$.
\end{proof}

\begin{proposition-app}\label{prop-DNlog-implies-DN-2}
  Let $\Delta$ be a filter that is DNlog for a clause $r$,
  $Q\lra_r T$ be a derivation step and
  $Q'$ be a query that is $\Delta$-more general than $Q$.
  Then, there exists a derivation step $Q'\lra_r T'$
  where $T'$ satisfies $\Delta$.
\end{proposition-app}
\begin{proof}
  Let $(\tau,\delta):=\Delta$ and $H\leftarrow c \diamond B :=r$.
  As $Q'$ is $\Delta$-more general than $Q$,
  $\Set{\restrict{Q}{\overline{\tau}}}\subseteq
  \Set{\restrict{Q'}{\overline{\tau}}}$.
  Moreover, as $Q\lra_r T$, by Lemma~\ref{lemma-operational-sem}
  we have $\Set{Q}\cap\Set{\query{H}{c}}\neq\emptyset$.
  So, by Lemma~\ref{lemma-useful2-theo-DNlog-iff-DN},
  $\Set{\restrict{Q}{\overline{\tau}}}\cap
  \Set{\restrict{\query{H}{c}}{\overline{\tau}}}\neq\emptyset.$
  Hence,
  \begin{equation}\label{prop-DNlog-implies-DN-2-eq3}
    \Set{\restrict{Q'}{\overline{\tau}}}\cap
    \Set{\restrict{\query{H}{c}}{\overline{\tau}}}\neq\emptyset~.
  \end{equation}

  Let $\query{p(\tilde{u}')}{d'}:=Q'$ and
  $r':=p(\tilde{X}')\leftarrow c'\diamond q(\tilde{Y}')$
  be a variant of $r$ variable disjoint with $Q'$.
  By Lemma~\ref{lemma-set-variant}, we have
  $\Set{\query{H}{c}}=\Set{\query{p(\tilde{X}')}{c'}}$ which implies,
  by Lemma~\ref{lemma-restriction}, that
  $\Set{\restrict{\query{H}{c}}{\overline{\tau}}}\subseteq
  \Set{\restrict{\query{p(\tilde{X}')}{c'}}{\overline{\tau}}}$
  \ie{}, by~(\ref{prop-DNlog-implies-DN-2-eq3}), that
  $\Set{\restrict{Q'}{\overline{\tau}}}\cap
  \Set{\restrict{\query{p(\tilde{X}')}{c'}}{\overline{\tau}}}\neq\emptyset$.
  Therefore, by Lemma~\ref{lemma-useful-theo-DNlog-iff-DN},
  there exists a valuation $v$ such that
  \begin{equation}\label{lemma-useful-theo-DNlog-iff-DN-eq00}
    \mathcal{D} \models_v
    (\restrict{\tilde{X}'}{\overline{\tau}(p)}=
    \restrict{\tilde{u}'}{\overline{\tau}(p)}
    \land c'\land d')\;.
  \end{equation}
  As $\Delta$ is DNlog for $r$, it is also DNlog for $r'$.
  Hence, if we let
  $\mathcal{Y}:=\restrict{\tilde{Y}'}{\tau(q)}\cup
  \mathit{local\_vars}(r')$, we have
  $\mathcal{D} \models_v
  c'\rightarrow \forall_{\restrict{\tilde{X}'}{\tau(p)}} \big[
  \sat{\restrict{\tilde{X}'}{\tau(p)}}{\delta(p)} \rightarrow
  \exists_{\mathcal{Y}} c'\big]$.
  As by~(\ref{lemma-useful-theo-DNlog-iff-DN-eq00})
  $\mathcal{D} \models_v c'$, then
  \begin{equation}\label{lemma-useful-theo-DNlog-iff-DN-eq01}
    \mathcal{D} \models_v
    \forall_{\restrict{\tilde{X}'}{\tau(p)}} \big[
    \sat{\restrict{\tilde{X}'}{\tau(p)}}{\delta(p)} \rightarrow
    \exists_{\mathcal{Y}} c' \big]\;.
  \end{equation}
  Let $v_1$ be the valuation defined as:
  \begin{itemize}
  \item for all variable $V\not\in\restrict{\tilde{X}'}{\tau(p)}$,
   $v_1(V)=v(V)$ and
  \item $v_1(\restrict{\tilde{X}'}{\tau(p)})=
    [\restrict{\tilde{u}'}{\tau(p)}]_v$.
  \end{itemize}
  Then by~(\ref{lemma-useful-theo-DNlog-iff-DN-eq01}) we have:
  \begin{equation}\label{lemma-useful-theo-DNlog-iff-DN-eq0}
    \mathcal{D} \models_{v_1}
    \sat{\restrict{\tilde{X}'}{\tau(p)}}{\delta(p)} \rightarrow
    \exists_{\mathcal{Y}} c'\;.
  \end{equation}
  Notice that
  $\restrict{p}{\tau}([\restrict{\tilde{u}'}{\tau(p)}]_v)\in
  \Set{\restrict{Q'}{\tau}}$ because,
  by~(\ref{lemma-useful-theo-DNlog-iff-DN-eq00}),
  $\mathcal{D} \models_v d'$.
  Moreover, as $Q'$ satisfies $\Delta$,
  $\Set{\restrict{Q'}{\tau}}\subseteq\Set{\delta(p)}$.
  Hence, $\restrict{p}{\tau}([\restrict{\tilde{u}'}{\tau(p)}]_v)\in
  \Set{\delta(p)}$.
  As, by definition of $v_1$,
  $[\restrict{\tilde{u}'}{\tau(p)}]_v = v_1(\restrict{\tilde{X}'}{\tau(p)})$
  with $v_1(\restrict{\tilde{X}'}{\tau(p)})=
  [\restrict{\tilde{X}'}{\tau(p)}]_{v_1}$
  (by definition of $[\cdot]_{v_1}$), we have
  $\restrict{p}{\tau}([\restrict{\tilde{X}'}{\tau(p)}]_{v_1})\in
  \Set{\delta(p)}$.
  So, by Lemma~\ref{lemma-sat-set},
  $\mathcal{D} \models_{v_1}
  \sat{\restrict{\tilde{X}'}{\tau(p)}}{\delta(p)}$.
  Hence, by~(\ref{lemma-useful-theo-DNlog-iff-DN-eq0}),
  $\mathcal{D} \models_{v_1} \exists_{\mathcal{Y}} c'$.
  Therefore, there exists a valuation $v_2$ such that:
  \begin{itemize}
  \item for all variable $V\not\in \mathcal{Y}$,
    $v_2(V)=v_1(V)$ and
  \item $\mathcal{D} \models_{v_2} c'$.
  \end{itemize}
  Notice that as $\Var(Q')\cap\Var(r')=
  \restrict{\tilde{X}'}{\overline{\tau}(p)}\cap
  (\restrict{\tilde{X}'}{\tau(p)}\cup\mathcal{Y})=\emptyset$,
  by definition of $v_2$ and $v_1$ we have
  $[d']_{v_2}=[d']_{v_1}=[d']_{v}$,
  $[\tilde{u}']_{v_2}=[\tilde{u}']_{v_1}=[\tilde{u}']_v$
  and $[\restrict{\tilde{X}'}{\overline{\tau}(p)}]_{v_2}=
  [\restrict{\tilde{X}'}{\overline{\tau}(p)}]_{v_1}=
  [\restrict{\tilde{X}'}{\overline{\tau}(p)}]_v$.
  So, by~(\ref{lemma-useful-theo-DNlog-iff-DN-eq00}), we have
  \begin{equation}\label{lemma-useful-theo-DNlog-iff-DN-eq5}
    \mathcal{D} \models_{v_2}
    (\restrict{\tilde{X}'}{\overline{\tau}(p)}=
    \restrict{\tilde{u}'}{\overline{\tau}(p)}
    \land c'\land d')\;.
  \end{equation}
  As $\restrict{\tilde{X}'}{\tau(p)}$ and
  $\mathcal{Y}$ are variable disjoint,  
  $[\restrict{\tilde{X}'}{\tau(p)}]_{v_2}=
  [\restrict{\tilde{X}'}{\tau(p)}]_{v_1}$
  with $[\restrict{\tilde{X}'}{\tau(p)}]_{v_1}=
  [\restrict{\tilde{u}'}{\tau(p)}]_v$
  by definition of $v_1$ and $[\restrict{\tilde{u}'}{\tau(p)}]_v=
  [\restrict{\tilde{u}'}{\tau(p)}]_{v_2}$.
  So, $\mathcal{D} \models_{v_2}
  (\restrict{\tilde{X}'}{\tau(p)}=\restrict{\tilde{u}'}{\tau(p)})$.
  Hence by~(\ref{lemma-useful-theo-DNlog-iff-DN-eq5})
  $\mathcal{D} \models_{v_2}(\tilde{X}'=\tilde{u}' \land c'\land d')$
  \ie{}
  $\mathit{solv} (\tilde{X}'=\tilde{u}' \land c'\land d')=\mathtt{true}$.
  Consequently, we have
  \[Q'\lra_r T' \quad\text{where}\quad
  T'=\query{q(\tilde{Y}')}{\tilde{X}'=\tilde{u}' \land c'\land d'}~.\]

  Let us prove that $T'$ satisfies $\Delta$ \ie{} that
  $\Set{\restrict{T'}{\tau}}\subseteq\Set{\delta(q)}$.
  Let $\restrict{q}{\tau}(\tilde{a})\in\Set{\restrict{T'}{\tau}}$.
  Then, there exists a valuation $w$ such that
  \begin{equation}\label{lemma-useful-theo-DNlog-iff-DN-eq7}
  \tilde{a}=[\restrict{\tilde{Y}'}{\tau(q)}]_w
  \quad
  \text{and}
  \quad
  \mathcal{D}\models_w \tilde{X}'=\tilde{u}' \land c'\land d'~.
  \end{equation}
  As $\Delta$ is DNlog for $r'$, we have
  $\mathcal{D}\models_w c'\rightarrow
  \sat{\restrict{\tilde{Y}'}{\tau(q)}}{\delta(q)}$.
  As $\mathcal{D}\models_w c'$
  (by~(\ref{lemma-useful-theo-DNlog-iff-DN-eq7})), then we have
  $\mathcal{D}\models_w \sat{\restrict{\tilde{Y}'}{\tau(q)}}{\delta(q)}$.
  So, by Lemma~\ref{lemma-sat-set},
  $\restrict{q}{\tau}([\restrict{\tilde{Y}'}{\tau(q)}]_w)\in\Set{\delta(q)}$.
  As $\tilde{a}=[\restrict{\tilde{Y}'}{\tau(q)}]_w$
  (by~(\ref{lemma-useful-theo-DNlog-iff-DN-eq7})),
  we have $\restrict{q}{\tau}(\tilde{a})\in\Set{\delta(q)}$.
\end{proof}

\begin{proposition-app}\label{prop-DNlog-implies-DN-3}
  Let $\Delta:=(\tau,\delta)$ be a filter that is DNlog for
  a clause $r$. Let $Q\lra_r T$ and $Q'\lra_r T'$ be some
  derivation steps such that $Q'$ is $\Delta$-more general
  than $Q$. Then, $\restrict{T'}{\overline{\tau}}$ is more
  general than $\restrict{T}{\overline{\tau}}$.
\end{proposition-app}
\begin{proof}
  Let $\query{p(\tilde{u})}{d} := Q$ and
  $\query{p(\tilde{u}')}{d'} := Q'$.
  Let $r_1 := p(\tilde{X})\la c\diamond q(\tilde{Y})$
  be the input clause in $Q\lra_r T$ and
  $r'_1 := p(\tilde{X}')\la c'\diamond q(\tilde{Y}')$
  that in $Q'\lra_r T'$. Then, we have
  \[T = \query{q(\tilde{Y})}{\tilde{X}=\tilde{u}\land c \land d}
  \quad\text{and}\quad
  T' = \query{q(\tilde{Y}')}{\tilde{X}'=\tilde{u}'\land c' \land d'}~.\]
  Let us prove that 
  $\Set{\restrict{T}{\overline{\tau}}}\subseteq
  \Set{\restrict{T'}{\overline{\tau}}}$. Let
  $\restrict{q}{\overline{\tau}}(\tilde{a})\in
  \Set{\restrict{T}{\overline{\tau}}}$. Then, there exists a valuation
  $v$ such that
  \begin{equation}\label{prop-DNlog-implies-DN-3-eq1}
    \tilde{a}=[\restrict{\tilde{Y}}{\overline{\tau}(q)}]_v
    \quad\text{and}\quad
    \mathcal{D}\models_v \tilde{X}=\tilde{u}\land c \land d~.
  \end{equation}
  So, $\mathcal{D}\models_v d$, hence 
  $\restrict{p}{\overline{\tau}}
  ([\restrict{\tilde{u}}{\overline{\tau}(p)}]_v) \in
  \Set{\restrict{Q}{\overline{\tau}}}$. As
  $\Set{\restrict{Q}{\overline{\tau}}}\subseteq
  \Set{\restrict{Q'}{\overline{\tau}}}$ (because $Q'$ is
  $\Delta$-more general than $Q$), we have
  $\restrict{p}{\overline{\tau}}
  ([\restrict{\tilde{u}}{\overline{\tau}(p)}]_v) \in
  \Set{\restrict{Q'}{\overline{\tau}}}$. So, there exists a valuation
  $v'_1$ such that
  \begin{equation}\label{prop-DNlog-implies-DN-3-eq2}
    [\restrict{\tilde{u}}{\overline{\tau}(p)}]_v =
    [\restrict{\tilde{u}'}{\overline{\tau}(p)}]_{v'_1}
    \quad\text{and}\quad
    \mathcal{D}\models_{v'_1} d'~.
  \end{equation}
  Notice that $r_1$ and $r'_1$ are variants, so $r_1=\gamma(r'_1)$
  for a renaming $\gamma$. As $Q'$ and
  $r'_1$ are variable disjoint (because $r'_1$ is the input clause in
  $Q'\lra_r T'$), there exists a valuation $v'$ such that:
  \begin{itemize}
  \item for all variable $V\in\Var(r'_1)$, $v'(V)=v(\gamma(V))$ and 
  \item for all variable $V\in\Var(Q')$, $v'(V)=v'_1(V)$. 
  \end{itemize}
  Then, we have
  \begin{itemize}
  \item $[\restrict{\tilde{X}'}{\overline{\tau}(p)}]_{v'}
    \mathop{=}\limits_{\text{def } v'}
    [\gamma(\restrict{\tilde{X}'}{\overline{\tau}(p)})]_v
    \mathop{=}\limits_{\text{def } \gamma}
    [\restrict{\tilde{X}}{\overline{\tau}(p)}]_v
    \mathop{=}\limits_{(\ref{prop-DNlog-implies-DN-3-eq1})}
    [\restrict{\tilde{u}}{\overline{\tau}(p)}]_v
    \mathop{=}\limits_{(\ref{prop-DNlog-implies-DN-3-eq2})}
    [\restrict{\tilde{u}'}{\overline{\tau}(p)}]_{v'_1}
    \mathop{=}\limits_{\text{def } v'}
    [\restrict{\tilde{u}'}{\overline{\tau}(p)}]_{v'}$
  \item $[c']_{v'} \mathop{=}\limits_{\text{def } v'}
    [\gamma(c')]_v \mathop{=}\limits_{\text{def } \gamma}
    [c]_v \mathop{=}\limits_{(\ref{prop-DNlog-implies-DN-3-eq1})} 1$
    and
    $[d']_{v'} \mathop{=}\limits_{\text{def } v'}
    [d']_{v'_1} \mathop{=}\limits_{(\ref{prop-DNlog-implies-DN-3-eq2})}1$
  \item $[\restrict{\tilde{Y}'}{\overline{\tau}(q)}]_{v'}
    \mathop{=}\limits_{\text{def } v'}
    [\gamma(\restrict{\tilde{Y}'}{\overline{\tau}(q)})]_v
    \mathop{=}\limits_{\text{def } \gamma}
    [\restrict{\tilde{Y}}{\overline{\tau}(q)}]_v
    \mathop{=}\limits_{(\ref{prop-DNlog-implies-DN-3-eq1})}
    \tilde{a}$.
  \end{itemize}
  Hence,
  \begin{equation}\label{prop-DNlog-implies-DN-3-eq3}
    \mathcal{D} \models_{v'}
    \restrict{\tilde{X}'}{\overline{\tau}(p)} =
    \restrict{\tilde{u}'}{\overline{\tau}(p)}
    \land c'\land d'
    \quad\text{and}\quad
    [\restrict{\tilde{Y}'}{\overline{\tau}(q)}]_{v'} = \tilde{a}~.
  \end{equation}
  As $\Delta$ is DNlog for $r$, then
  it is DNlog for $r'_1$. Consequently, if we let
  $\mathcal{Y}:=\restrict{\tilde{Y}'}{\tau(q)}\cup
  \mathit{local\_vars}(r'_1)$, we have
  $\mathcal{D} \models_{v'}
  c'\rightarrow \forall_{\restrict{\tilde{X}'}{\tau(p)}} \big[
  \sat{\restrict{\tilde{X}'}{\tau(p)}}{\delta(p)} \rightarrow
  \exists_{\mathcal{Y}} c'\big]$.
  As, by~(\ref{prop-DNlog-implies-DN-3-eq3}),
  $\mathcal{D}\models_{v'} c'$, we have
  $\mathcal{D} \models_{v'}
  \forall_{\restrict{\tilde{X}'}{\tau(p)}} \big[
  \sat{\restrict{\tilde{X}'}{\tau(p)}}{\delta(p)} \rightarrow
  \exists_{\mathcal{Y}} c' \big].$
  Let $w'$ be the valuation defined as:
  \begin{itemize}
  \item for all variable $V\not\in\restrict{\tilde{X}'}{\tau(p)}$,
   $w'(V)=v'(V)$ and
  \item $w'(\restrict{\tilde{X}'}{\tau(p)})=
    [\restrict{\tilde{u}'}{\tau(p)}]_{v'}$.
  \end{itemize}
  Then,
  \begin{equation}\label{prop-DNlog-implies-DN-3-eq4}
    \mathcal{D} \models_{w'}
    \sat{\restrict{\tilde{X}'}{\tau(p)}}{\delta(p)} \rightarrow
    \exists_{\mathcal{Y}} c'\;.
  \end{equation}
  Notice that
  $\restrict{p}{\tau}([\restrict{\tilde{u}'}{\tau(p)}]_{v'}) \in
  \Set{\restrict{Q'}{\tau}}$ because $\mathcal{D}\models_{v'} d'$
  by~(\ref{prop-DNlog-implies-DN-3-eq3}). As
  $[\restrict{\tilde{u}'}{\tau(p)}]_{v'}=
  [\restrict{\tilde{X}'}{\tau(p)}]_{w'}$ (by definition of $w'$),
  then $\restrict{p}{\tau}([\restrict{\tilde{X}'}{\tau(p)}]_{w'}) \in
  \Set{\restrict{Q'}{\tau}}$. As $Q'$ is $\Delta$-more general than
  $Q$, we have $\Set{\restrict{Q'}{\tau}}\subseteq \Set{\delta(p)}$.
  Hence, $\restrict{p}{\tau}([\restrict{\tilde{X}'}{\tau(p)}]_{w'}) \in
  \Set{\delta(p)}$. So, by Lemma~\ref{lemma-sat-set},
  $\mathcal{D}\models_{w'}\sat{\restrict{\tilde{X}'}{\tau(p)}}{\delta(p)}$.
  Therefore, we have $\mathcal{D} \models_{w'} \exists_{\mathcal{Y}} c'$
  by~(\ref{prop-DNlog-implies-DN-3-eq4}). Hence, there exists
  a valuation $w'_1$ such that:
  \begin{itemize}
  \item for all variable $V\not\in\mathcal{Y}$, $w'_1(V)=w'(V)$ and
  \item $\mathcal{D} \models_{w'_1} c'$.
  \end{itemize}
  Then, as $\restrict{\tilde{X}'}{\tau(p)} \cap \mathcal{Y} =
  \restrict{\tilde{X}'}{\overline{\tau}(p)}
  \cap (\mathcal{Y} \cup \restrict{\tilde{X}'}{\tau(p)}) =
  \Var(Q')\cap\Var(r'_1) = \emptyset$,
  we have
  \begin{itemize}
  \item $[\restrict{\tilde{X}'}{\tau(p)}]_{w'_1}
    \mathop{=}\limits_{\text{def } w'_1}
    [\restrict{\tilde{X}'}{\tau(p)}]_{w'}
    \mathop{=}\limits_{\text{def } w'}
    [\restrict{\tilde{u}'}{\tau(p)}]_{v'}
    \mathop{=}\limits_{\text{def } w' + \text{def }w'_1}
    [\restrict{\tilde{u}'}{\tau(p)}]_{w'_1}$,
  \item $[\restrict{\tilde{X}'}{\overline{\tau}(p)}]_{w'_1}
    \mathop{=}\limits_{\text{def } w'_1}
    [\restrict{\tilde{X}'}{\overline{\tau}(p)}]_{w'}
    \mathop{=}\limits_{\text{def } w'}
    [\restrict{\tilde{X}'}{\overline{\tau}(p)}]_{v'}
    \mathop{=}\limits_{(\ref{prop-DNlog-implies-DN-3-eq3})}
    [\restrict{\tilde{u}'}{\overline{\tau}(p)}]_{v'}
    \mathop{=}\limits_{\text{def } w' + \text{def }w'_1}
    [\restrict{\tilde{u}'}{\overline{\tau}(p)}]_{w'_1}$.
  \item $[d']_{w'_1}
    \mathop{=}\limits_{\text{def } w'_1 + \text{def }w'}
    [d']_{v'}
    \mathop{=}\limits_{(\ref{prop-DNlog-implies-DN-3-eq3})} 1$.
  \end{itemize}
  So, $\mathcal{D}\models_{w'_1} \tilde{X}'=\tilde{u}'\land c' \land d'$.
  Moreover, $[\restrict{\tilde{Y}'}{\overline{\tau}(q)}]_{w'_1} =
  [\restrict{\tilde{Y}'}{\overline{\tau}(q)}]_{w'} =
  [\restrict{\tilde{Y}'}{\overline{\tau}(q)}]_{v'}$ with
  $[\restrict{\tilde{Y}'}{\overline{\tau}(q)}]_{v'} = \tilde{a}$
  by~(\ref{prop-DNlog-implies-DN-3-eq3}).
  Consequently, $\restrict{q}{\overline{\tau}}(\tilde{a})\in
  \Set{\restrict{T'}{\overline{\tau}}}$.
\end{proof}

\subsection{-- Theorem~\ref{prop-DN-implies-DNlog2}}
Suppose that for all atoms $A$ whose arguments are elements of $D$,
there exists a query $Q$ such that $\Set{Q}=\{A\}$. Given a clause
$r:=p(\tilde{X})\leftarrow c \diamond q(\tilde{Y})$ and a filter
$\Delta:=(\tau,\delta)$ that is DN for $r$, we have to prove that
$\Delta$ is DNlog for $r$. By Definition~\ref{def-log-DN},
we have to establish that
\begin{itemize}
\item $\mathrm{DNlog1}(\Delta,r) := \big(c \rightarrow
  \forall_{\restrict{\tilde{X}}{\tau(p)}} \big[
  \sat{\restrict{\tilde{X}}{\tau(p)}}{\delta(p)} \rightarrow
  \exists_{\mathcal{Y}} c\big]\big)$ and
\item $\mathrm{DNlog2}(\Delta,r) := \big(c\rightarrow
  \sat{\restrict{\tilde{Y}}{\tau(q)}}{\delta(q)}\big)$
\end{itemize}
hold.
Proposition~\ref{DN-implies-DNlog-1} below establishes that
$\mathrm{DNlog1}(\Delta,r)$ is true and
Proposition~\ref{DN-implies-DNlog-1} below establishes that
$\mathrm{DNlog2}(\Delta,r)$ is true.

%
\begin{proposition-app}\label{DN-implies-DNlog-1}
  Assume that the following holds: for each atom $A$
  whose arguments are elements of $D$,
  there exists a query $Q$ such that
  $\Set{Q}=\{A\}$.
  Let $\Delta$ be a filter that is DN for a clause $r$.
  Then, $\mathcal{D} \models \mathrm{DNlog1}(\Delta,r)$.
\end{proposition-app}
\begin{proof}
  We let $\Delta:=(\tau,\delta)$ and
  $r := p(\tilde{X}) \la c \diamond q(\tilde{Y})$.

  Let $v$ be a valuation. Suppose that
  \begin{equation}\label{prop-DN-implies-DNlog2-eq1}
    \mathcal{D}\models_v c\;.
  \end{equation}
  Let $v'$ be a valuation such that for all variable
  $V\not\in \restrict{\tilde{X}}{\tau(p)}$, $v'(V)=v(V)$.
  Suppose that
  \begin{equation}\label{prop-DN-implies-DNlog2-eq2}
    \mathcal{D}\models_{v'}
    \sat{\restrict{\tilde{X}}{\tau(p)}}{\delta(p)}\;.
  \end{equation}
  Notice that for all variable $V$, $[V]_v\in D$
  and $[V]_{v'}\in D$.
  So, there exists a query $Q$
  such that $\Set{Q}=\{p([\tilde{X}]_v)\}$
  and a query  $Q'$
  such that $\Set{\restrict{Q'}{\tau}}=
  \{\restrict{p}{\tau}([\restrict{\tilde{X}}{\tau(p)}]_{v'})\}$
  and $\Set{\restrict{Q'}{\overline{\tau}}}=
  \{\restrict{p}{\overline{\tau}}
  ([\restrict{\tilde{X}}{\overline{\tau}(p)}]_v)\}$.
  Let us prove that $\mathcal{D}\models_{v'} \exists_{\mathcal{Y}} c$.
  We have:
  \begin{itemize}
  \item $\Set{\restrict{Q}{\overline{\tau}}}=
    \{\restrict{p}{\overline{\tau}}
    ([\restrict{\tilde{X}}{\overline{\tau}(p)}]_v)\}=
    \Set{\restrict{Q'}{\overline{\tau}}}$
    and
  \item $\restrict{p}{\tau}
    ([\restrict{\tilde{X}}{\tau(p)}]_{v'})\in\Set{\delta(p)}$
    because $\mathcal{D} \models_{v'}
    \sat{\restrict{\tilde{X}}{\tau(p)}}{\delta(p)}$ and    
    by Lemma~\ref{lemma-sat-set}.
    So, as $\Set{\restrict{Q'}{\tau}}=
    \{\restrict{p}{\tau}([\restrict{\tilde{X}}{\tau(p)}]_{v'})\}$, we have
    $\Set{\restrict{Q'}{\tau}}\subseteq\Set{\delta(p)}$.
  \end{itemize} 
  Consequently, $Q'$ is $\Delta$-more general than $Q$. Moreover,
  as $\mathcal{D} \models_v c$, we have
  $p([\tilde{X}]_v)\in\Set{\query{p(\tilde{X})}{c}}$. As
  $\Set{Q}=\{p([\tilde{X}]_v)\}$, this implies that 
  $\Set{Q}\cap \Set{\query{p(\tilde{X})}{c}} \neq \emptyset$.
  Hence, by Lemma~\ref{lemma-operational-sem}, there exists
  a derivation step of the form $Q\lra_r T$. Let
  $r_1:=p(\tilde{U}) \leftarrow c_1\diamond q(\tilde{V})$
  be the input clause in this derivation step. Then, 
  if we let $Q:=\query{p(\tilde{t})}{d}$, we have
  \[T=\query{q(\tilde{V})}{\tilde{U}=\tilde{t}\land c_1\land d}~.\]
  As $Q'$ is $\Delta$-more general than $Q$ and $\Delta$ is DN
  for $r$, there exists a query $T'$ such that
  $Q'\lra_r T'$ and $T'$ is $\Delta$-more general than $T$.
  Let $r'_1:=p(\tilde{U}') \leftarrow c'_1\diamond q(\tilde{V}')$
  be the input clause in $Q'\lra_r T'$. Then,
  if we let $Q':=\query{p(\tilde{t}')}{d'}$, we have
  \[T'=\query{q(\tilde{V}')}{\tilde{U}'=\tilde{t}'\land c'_1\land d'}~.\]
  As $r_1$ is a variant of $r$, there exists a renaming $\gamma$
  such that $r=\gamma(r_1)$.
  Let $v_1$ be the valuation defined as:
  \begin{itemize}
  \item for all variable $V\in\Var(r_1)$, $v_1(V)=v(\gamma(V))$ and
  \item for all variable $V\not\in\Var(r_1)$, $v_1(V)=v(V)$.
  \end{itemize}
  As $\Set{Q}=\{p([\tilde{X}]_v)\}$, there exists a valuation $v_Q$
  such that
  \begin{equation}\label{prop-DN-implies-DNlog2-eq2-1}
    [\tilde{t}]_{v_Q} = [\tilde{X}]_v \quad\text{and}\quad
    \mathcal{D} \models_{v_Q} d~.
  \end{equation}
  Let $v_2$ be the valuation defined as:
  \begin{itemize}
  \item for all variable $V\in\Var(Q)$, $v_2(V)=v_Q(V)$ and
  \item for all variable $V\not\in\Var(Q)$, $v_2(V)=v_1(V)$.
  \end{itemize}
  As $\Var(Q)\cap\Var(r_1)=\emptyset$ (because $r_1$
  is the input clause in $Q\lra_r T$), we have
  \[[c_1]_{v_2}
  \mathop{=}\limits_{\text{def } v_2}
  [c_1]_{v_1}
  \mathop{=}\limits_{\text{def } v_1}
  [\gamma(c_1)]_v
  \mathop{=}\limits_{\text{def } \gamma}
  [c]_v
  \mathop{=}\limits_{(\ref{prop-DN-implies-DNlog2-eq1})}
  1 \quad \text{and}\]
  \[[\tilde{U}]_{v_2}
  \mathop{=}\limits_{\text{def } v_2}
  [\tilde{U}]_{v_1}
  \mathop{=}\limits_{\text{def } v_1}
  [\gamma(\tilde{U})]_v
  \mathop{=}\limits_{\text{def } \gamma}
  [\tilde{X}]_v
  \mathop{=}\limits_{(\ref{prop-DN-implies-DNlog2-eq2-1})}
  [\tilde{t}]_{v_Q}
  \mathop{=}\limits_{\text{def } v_2}
  [\tilde{t}]_{v_2}\;.\]
  Moreover,
  $[d]_{v_2}
  \mathop{=}\limits_{\text{def } v_2}
  [d]_{v_Q}
  \mathop{=}\limits_{(\ref{prop-DN-implies-DNlog2-eq2-1})}
  1$. Consequently,
  \[\mathcal{D}\models_{v_2} \tilde{U}=\tilde{t}\land c_1 \land d~.\]
  So, $\restrict{q}{\overline{\tau}}
  ([\restrict{\tilde{V}}{\overline{\tau}(q)}]_{v_2})\in
  \Set{\restrict{T}{\overline{\tau}}}$. As
  $[\restrict{\tilde{V}}{\overline{\tau}(q)}]_{v_2}
  \mathop{=}\limits_{\text{def } v_2}
  [\restrict{\tilde{V}}{\overline{\tau}(q)}]_{v_1}
  \mathop{=}\limits_{\text{def } v_1}
  [\gamma(\restrict{\tilde{V}}{\overline{\tau}(q)})]_v
  \mathop{=}\limits_{\text{def } \gamma}
  [\restrict{\tilde{Y}}{\overline{\tau}(q)}]_v$ we have
  $\restrict{q}{\overline{\tau}}
  ([\restrict{\tilde{Y}}{\overline{\tau}(q)}]_v)\in
  \Set{\restrict{T}{\overline{\tau}}}$. Moreover, as
  $\Set{\restrict{T}{\overline{\tau}}}\subseteq
  \Set{\restrict{T'}{\overline{\tau}}}$
  (because $T'$ is $\Delta$-more general than $T$),
  $\restrict{q}{\overline{\tau}}
  ([\restrict{\tilde{Y}}{\overline{\tau}(q)}]_v)\in
  \Set{\restrict{T'}{\overline{\tau}}}$. Consequently,
  there exists a valuation $v'_1$ such that
  \begin{equation}\label{lemma2-prop-DN-implies-DNlog2-eq1}
    [\restrict{\tilde{V}'}{\overline{\tau}(q)}]_{v'_1} =
    [\restrict{\tilde{Y}}{\overline{\tau}(q)}]_v \quad\text{and}\quad
    \mathcal{D}\models_{v'_1} \tilde{U}' = \tilde{t}' \land c'_1 \land d'~.
  \end{equation}
  As $r'_1$ is a variant of $r$, there exists a renaming $\gamma'$
  such that $r'_1=\gamma'(r)$.
  Let $w$ be a valuation such that for all variable
  $V\in\Var(r)$, $w(V)=v'_1(\gamma'(V))$.
  Then, $[c]_w = [\gamma'(c)]_{v'_1} = [c'_1]_{v'_1}
  \mathop{=}\limits_{(\ref{lemma2-prop-DN-implies-DNlog2-eq1})}
  1$, so
  \begin{equation}\label{lemma2-prop-DN-implies-DNlog2-eq3}
    \mathcal{D}\models_w c~.
  \end{equation}
  Notice that:
  \begin{itemize}
  \item $[\restrict{\tilde{X}}{\tau(p)}]_w
    \mathop{=}\limits_{\text{def } w}
    [\gamma'(\restrict{\tilde{X}}{\tau(p)})]_{v'_1}
    \mathop{=}\limits_{\text{def } \gamma'}
    [\restrict{\tilde{U}'}{\tau(p)}]_{v'_1}
    \mathop{=}\limits_{(\ref{lemma2-prop-DN-implies-DNlog2-eq1})}
    [\restrict{\tilde{t}'}{\tau(p)}]_{v'_1} =
    [\restrict{\tilde{X}}{\tau(p)}]_{v'}$ because,
    as $\mathcal{D} \models_{v'_1} d'$
    by~(\ref{lemma2-prop-DN-implies-DNlog2-eq1}),
    we have $\restrict{p}{\tau}
    ([\restrict{\tilde{t}'}{\tau(p)}]_{v'_1})\in\Set{\restrict{Q'}{\tau}} =
    \{\restrict{p}{\tau}([\restrict{\tilde{X}}{\tau(p)}]_{v'})\}$;
  \item $[\restrict{\tilde{X}}{\overline{\tau}(p)}]_w
    \mathop{=}\limits_{\text{def } w}
    [\gamma'(\restrict{\tilde{X}}{\overline{\tau}(p)})]_{v'_1}
    \mathop{=}\limits_{\text{def } \gamma'}
    [\restrict{\tilde{U}'}{\overline{\tau}(p)}]_{v'_1}
    \mathop{=}\limits_{(\ref{lemma2-prop-DN-implies-DNlog2-eq1})}
    [\restrict{\tilde{t}'}{\overline{\tau}(p)}]_{v'_1} =
    [\restrict{\tilde{X}}{\overline{\tau}(p)}]_v$ because,
    as $\mathcal{D} \models_{v'_1} d'$
    by~(\ref{lemma2-prop-DN-implies-DNlog2-eq1}),
    we have $\restrict{p}{\overline{\tau}}
    ([\restrict{\tilde{t}'}{\overline{\tau}(p)}]_{v'_1})
    \in\Set{\restrict{Q'}{\overline{\tau}}} =
    \{\restrict{p}{\overline{\tau}}
    ([\restrict{\tilde{X}}{\overline{\tau}(p)}]_v)\}$;
    moreover, by definition of $v'$,
    $[\restrict{\tilde{X}}{\overline{\tau}(p)}]_v
    =
    [\restrict{\tilde{X}}{\overline{\tau}(p)}]_{v'}$
    because
    $\restrict{\tilde{X}}{\overline{\tau}(p)}\cap
    \restrict{\tilde{X}}{\tau(p)}=\emptyset$;
  \item $[\restrict{\tilde{Y}}{\overline{\tau}(q)}]_w
    \mathop{=}\limits_{\text{def } w}
    [\gamma'(\restrict{\tilde{Y}}{\overline{\tau}(q)})]_{v'_1}
    \mathop{=}\limits_{\text{def } \gamma'}
    [\restrict{\tilde{V}'}{\overline{\tau}(q)}]_{v'_1}
    \mathop{=}\limits_{(\ref{lemma2-prop-DN-implies-DNlog2-eq1})}
    [\restrict{\tilde{Y}}{\overline{\tau}(q)}]_v
    \mathop{=}\limits_{\text{def } v'}
    [\restrict{\tilde{Y}}{\overline{\tau}(q)}]_{v'}$
    because we have that
    $\restrict{\tilde{Y}}{\overline{\tau}(q)}\cap
    \restrict{\tilde{X}}{\tau(p)}=\emptyset$.
  \end{itemize}
  Consequently, as
  $\Var(c)\setminus\mathcal{Y}\subseteq
  \restrict{\tilde{X}}{\tau(p)}\cup
  \restrict{\tilde{X}}{\overline{\tau}(p)}\cup
  \restrict{\tilde{Y}}{\overline{\tau}(q)}$, we have:
  \begin{equation}\label{lemma2-prop-DN-implies-DNlog2-eq4}
    \text{for all } V\in\Var(c)\setminus\mathcal{Y},\
    w(V)=v'(V)\;.
  \end{equation}
  Let $w_1$ be the valuation defined as:
  \begin{itemize}
  \item for all variable $V\not\in\mathcal{Y}$,
    $w_1(V)=v'(V)$ and
  \item for all variable $V\in\mathcal{Y}$,
    $w_1(V)=w(V)$.
  \end{itemize}
  Then, for all variable $V\in\Var(c)$, if $V\in\mathcal{Y}$ then
  $w_1(V)\mathop{=}\limits_{\text{def } w_1}w(V)$ and
  if $V\not\in\mathcal{Y}$
  then $w_1(V)\mathop{=}\limits_{\text{def } w_1}v'(V)
  \mathop{=}\limits_{(\ref{lemma2-prop-DN-implies-DNlog2-eq4})}w(V)$.
  Consequently,
  $[c]_{w_1}=[c]_w
  \mathop{=}\limits_{(\ref{lemma2-prop-DN-implies-DNlog2-eq3})}
  1$.
  So, $\mathcal{D}\models_{w_1} c$
  which implies, by definition of $w_1$, that
  $\mathcal{D}\models_{v'} \exists_{\mathcal{Y}} c$.

  Hence, as we supposed~(\ref{prop-DN-implies-DNlog2-eq2}),
  we have 
  $\mathcal{D}\models_{v'}
  \sat{\restrict{\tilde{X}}{\tau(p)}}{\delta(p)} \rightarrow
  \exists_{\mathcal{Y}} c$.
  Therefore, as $v'$ denotes any valuation such that
  $v'(V)=v(V)$ for all variable
  $V\not\in\restrict{\tilde{X}}{\tau(p)}$,
  we get
  $\mathcal{D}\models_{v}
  \forall_{\restrict{\tilde{X}}{\tau(p)}} \big[
  \sat{\restrict{\tilde{X}}{\tau(p)}}{\delta(p)} \rightarrow
  \exists_{\mathcal{Y}} c \big]$.
  As we supposed~(\ref{prop-DN-implies-DNlog2-eq1}),
  we deduce that
  $\mathcal{D}\models_{v}
  c \rightarrow \forall_{\restrict{\tilde{X}}{\tau(p)}} \big[
  \sat{\restrict{\tilde{X}}{\tau(p)}}{\delta(p)} \rightarrow
  \exists_{\mathcal{Y}} c \big]$ where
  $v$ denotes any valuation. Hence the result.
\end{proof}

%
\begin{proposition-app}\label{DN-implies-DNlog-2}
  Let $\Delta$ be a filter that is DN for a clause $r$.
  Then, $\mathcal{D} \models \mathrm{DNlog2}(\Delta,r)$.
\end{proposition-app}
\begin{proof}
  We let $\Delta:=(\tau,\delta)$ and
  $r := p(\tilde{X}) \la c \diamond q(\tilde{Y})$.
  
  By Lemma~\ref{lemma-useful-loop}, there exists a derivation
  step
  \[\query{p(\tilde{X})}{c}\lra_r Q
  \quad\text{where}\quad
  \Set{Q}=\Set{\query{q(\tilde{Y})}{c}}~.\]
  Then, as $\Delta$ is DN for $r$, $Q$ satisfies $\Delta$ \ie{}
  $\Set{\restrict{Q}{\tau}}\subseteq\Set{\delta(q)}$.
  Moreover, as
  $\Set{\query{q(\tilde{Y})}{c}}\subseteq\Set{Q}$,
  by Lemma~\ref{lemma-restriction}
  $\Set{\restrict{\query{q(\tilde{Y})}{c}}{\tau}}
  \subseteq\Set{\restrict{Q}{\tau}}$.
  So,
  \begin{equation}\label{DN-implies-DNlog-2-eq1}
    \Set{\restrict{\query{q(\tilde{Y})}{c}}{\tau}}
    \subseteq\Set{\delta(q)}~.
  \end{equation}
  Let $v$ be a valuation. Suppose that
  \begin{equation}\label{DN-implies-DNlog-2-eq2}
    \mathcal{D}\models_v c~.
  \end{equation}
  Notice that $\restrict{q}{\tau}([\restrict{\tilde{Y}}{\tau(q)}]_v)
  \in \Set{\restrict{\query{q(\tilde{Y})}{c}}{\tau}}$.
  So, by~(\ref{DN-implies-DNlog-2-eq1}),
  $\restrict{q}{\tau}([\restrict{\tilde{Y}}{\tau(q)}]_v)
  \in \Set{\delta(q)}$.
  Therefore, by Lemma~\ref{lemma-sat-set}, we have
  $\mathcal{D}\models_v \sat{\restrict{\tilde{Y}}{\tau(q)}}{\delta(q)}$.
  As we supposed~(\ref{DN-implies-DNlog-2-eq2}), then we have
  $\mathcal{D}\models_v c \rightarrow 
  \sat{\restrict{\tilde{Y}}{\tau(q)}}{\delta(q)}$ where $v$
  denotes any valuation. Hence the result.
\end{proof}

\end{document}